\documentclass[11pt]{article}
\usepackage[utf8]{inputenc}

\usepackage{varioref}
\usepackage{amsmath,amsfonts,amsthm,mathrsfs}
\usepackage{mathabx}
\usepackage[normalem]{ulem}
\usepackage{multirow,color,graphics}
\usepackage{amsmath}
\usepackage{bbold}
\usepackage{amsfonts}
\usepackage{amssymb}
\usepackage{jheppub}
\usepackage{enumerate}
\usepackage{fancyvrb}
\usepackage{verbatim}
\usepackage{wrapfig}
\usepackage{appendix}
\usepackage{amstext}
\usepackage{amssymb}
\usepackage{graphicx}
\usepackage{color}
\usepackage{varioref}
\usepackage{multirow,graphics}
\usepackage{epstopdf}
\newcommand{\nn}{\nonumber}

\numberwithin{equation}{section}

\def\sym{{\rm sym}_{\{\theta_i\}}}
\def\[{\left[}
\def\]{\right]}
\def\({\left(}
\def\){\right)}
\def\<{\left<}
\def\>{\right>}
\def\d{\partial}

    \newcommand{\beq}{\begin{equation}}
    \newcommand{\eeq}{\end{equation}}
    \newcommand\beqa{\begin{eqnarray}}
    \newcommand\eeqa{\end{eqnarray}}
\newcommand\bea{\begin{array}}
\newcommand\eea{\end{array}}

\newcommand{\bQ}{{\bf Q}}

\newcommand{\la}[1]{\label{#1}}
\newcommand{\eq}[1]{(\ref{#1})}

    \def\bQ{{\bf Q}}

\definecolor{cadmiumgreen}{rgb}{0.0, 0.42, 0.24}

\makeatletter
     \@ifundefined{usebibtex}{} {}
\makeatother

    \def\bQ{{\bf Q}}

\newcommand{\ket}[1]{| #1 \rangle}
\newcommand{\cN}{\mathcal{N}}

\renewcommand{\<}{\langle} 
\renewcommand{\>}{\rangle} 
\renewcommand{\sl}{\mathfrak{sl}}

\title{
Separation of variables and scalar products at any rank
}

\author[a]{~~Andrea Cavagli\`a,}
\author[a,b]{~~Nikolay Gromov,}

\author[c,1]{~~Fedor Levkovich-Maslyuk\note{Also at Institute for Information Transmission Problems, Moscow 127994, Russia}}

\affiliation[a]{Mathematics Department, King's College London,
The Strand, London WC2R 2LS, UK}
\affiliation[b]{St.Petersburg INP, Gatchina, 188 300, St.Petersburg,
  Russia}

\affiliation[c]{Departement de Physique, Ecole Normale Superieure / PSL Research University, CNRS, 24 rue Lhomond, 75005 Paris, France }

\emailAdd{andrea.cavaglia$\bullet$kcl.ac.uk}
\emailAdd{nikgromov$\bullet$gmail.com}
\emailAdd{fedor.levkovich$\bullet$gmail.com} 

\abstract{
Separation of variables (SoV) is a special property of integrable models which ensures that the wavefunction has a very simple factorised form in a specially designed basis. 
Even though the factorisation of the wavefunction was recently established for higher rank models by two of the authors and G.~Sizov, 
the measure for the scalar product was not known beyond the case of  rank one symmetry. In this paper we
show how this measure can be found, bypassing an explicit SoV construction. 
A key new  observation is that the measure  for spin chains in a highest-weight infinite dimensional representation of $\sl(N)$ couples Q-functions at different nesting levels in a non-symmetric fashion. We also managed to express a large number of form factors as ratios of determinants in our new approach. 
We expect our method to be applicable in a much wider setup including the problem of computing correlators in integrable CFTs such as the fishnet theory, ${\cal N}=4$ SYM and the ABJM model.
}

\begin{document}

\maketitle

\section{Introduction}

Integrability provides powerful methods to study certain quantum systems at the nonperturbative level. 
The integrable models share many universal features and the underlying mathematical structures which to a great extent depend on the global symmetries of the model.  
While a bit counterintuitive, in fact  the integrable structure becomes increasingly more complex for systems with a larger symmetry. 

The observables that are most easily accessible to integrability are the eigenvalues of the Hamiltonian and other integrals of motion. 
 However, it is usually much more difficult to extract information on the energy eigenvectors, as well as on more complicated observables such as form factors and correlators. 
A very powerful method that allows one to make advance in this direction was pioneered by Sklyanin in \cite{Sklyanin:1984sb,Sklyanin:1987ih,Sklyanin:1991ss,Sklyanin:1995bm}  and is known as the Separation of Variables (SoV). It is based on the fact that in  integrable systems the wave functions $| \Psi \rangle $ for eigenstates of the integrals of motion (also known as Bethe states) are expected to factorize completely in a suitable system of coordinates. The most elementary example is the wavefunction for the hydrogen atom which factorizes in spherical coordinates. 

In the case of a spin chain with $L$ sites and rank-one $\sl(2)$ symmetry
one aims to find a basis for the Hilbert space parametrized by a set of $L$ separated variables, $\langle {\bf x}|$, labelled by  a set of $L$ real numbers ${\bf x}=\left\{ x_i \right\}_{i=1}^L$, such that the Bethe state becomes a product, 
 \beq\label{eq:fact1}
\Psi({\bf x})\equiv  \langle{\bf x} | \Psi \rangle = \prod_{i=1}^L Q(x_i) \;.
 \eeq
In most realizations of SoV, the one-particle factors coincide precisely with the Q-function, which is a fundamental object in quantum integrability directly related to the solution of the spectral problem. The Q-function is  fixed by the Baxter TQ relation, a finite difference equation of order equal to the rank of the symmetry group, which in the SoV framework can be interpreted as equivalent to the Schr\"odinger equation. 
The explicit form of the change of coordinates to the SoV basis is  quite complicated, however, in many important cases one could reformulate the problem directly in the SoV coordinates. It was observed that for a number of important observables the result written in the SoV basis is surprisingly simple \cite{Korepin:1982gg,Kazama:2013rya,Niccoli:2012ci,Levy-Bencheton:2015mia,Niccoli:2014sfa,Jiang:2015lda,Kitanine:2016pvg,Kitanine:2015jna,Kitanine:2014swa}.
Crucially, one can define the scalar product bypassing the original physical basis directly in SoV.
The scalar product involves the so-called Sklyanin's measure $M({\bf x})$ such that
$\langle  \Psi_A  |  \Psi_B \rangle=\int d^L {\bf x}\;\Psi_A({\bf x})M({\bf x})
\Psi_B({\bf x})
$. In particular for the Bethe states it becomes
 \beq\label{eq:mu0}
\langle  \textcolor{blue}{\Psi_A } |  \textcolor{cadmiumgreen}{\Psi_B }\rangle  \!=\! \int  d^L {\bf x}\, \(\textcolor{blue}{ \underbrace{\prod_{i=1}^L Q^{(A)}(x_i) }_{\text{state $A$} }} \)  \textcolor{red}{\underbrace{ M({\bf x} )}_{\text{measure}} } \( \textcolor{cadmiumgreen}{ \underbrace{\prod_{i=1}^L Q^{(B)}(x_i) }_{\text{state $B$}}}\) \;.
 \eeq
As emphasised by the colours in the formula, the two states are represented by the respective factorized wavefunctions. The scalar product is implemented by integration over the values of the separated variables with the measure $M({\bf x })$, which is independent of the states.

 Of course, different eigenstates of Hermitian integrals of motion are orthogonal, therefore the integral in (\ref{eq:mu0}) should vanish for any two different Bethe states $\langle\Psi_A|$ and $|\Psi_B\rangle$. One can, in fact, reverse the logic and derive the measure (up to a constant factor) from the orthogonality of the Bethe states. In this paper we use this fact as an inspiration for the generalisation to higher rank symmetries. 
 
 Originally, for $\sl(2)$  spin chains  the basis $\ket{ {\bf x} }$ was constructed explicitly  and the measure derived rigorously in \cite{Derkachov:2001yn,Derkachov:2002wz}. It was shown to take the form  of a determinant of a $L\times L$ matrix\footnote{
 This formula is valid for the most general case of inhomogeneous spin chain with twisted boundary conditions.
 The reduction to the untwisted case can be obtained by carefully taking the corresponding limit. }\ \footnote{For compact spin chains we have sums instead of integrals and the corresponding measure was derived in e.g. \cite{Kazama:2013rya} (see also \cite{Niccoli:2012vq,Jiang:2015lda,Gromov:2016itr}).}  \beq\label{eq:musl2}
    M({\bf x}) = \sym \begin{vmatrix} \left(\frac{x_i^{j-1} }{1 + e^{2 \pi (x_i - \theta_i ) } } \right) \end{vmatrix}_{1\leq i,j\leq L} \;, 
  \eeq
  where  $\theta_i$ are $L$ distinct  inhomogeneities, and  ``$\text{sym}$'' denotes symmetrization of the determinant w.r.t. the inhomogeneities. This operation makes the expression \eq{eq:musl2} completely symmetric w.r.t. the   variables $x_i$, and does not affect the integral defining the scalar product.
 
 For models with higher rank symmetries, SoV methods have so far not been understood to the same extent as we described above. In fact, problems with obtaining the measure were anticipated recently in \cite{Martin:2015eea}. At the same time, extra motivation to explore this direction comes from  string theory and integrability observed in ${\cal N}=4$ super Yang-Mills, which has a much more complicated $\mathfrak{psu}(2,2|4)$ symmetry \cite{Beisert:2010jr}.

It was essentially conjectured in the original papers of Sklyanin \cite{Sklyanin:1992eu,Sklyanin:1992sm} how to construct the SoV basis $|x_i\rangle$ in the first higher rank case, i.e. for $\sl(3)$. These results were extended to $\sl(N)$ by Smirnov \cite{Smirnov2001} following the classical case \cite{Scott:1994dz,Ge95} (see also  \cite{Chervov:2006xk,Chervov:2007bb}). However, for a long time there was no precise indication of how Bethe states can be written in the separated coordinates and what are the corresponding factors.

One of the obvious difficulties  in generalizing (\ref{eq:mu0}) to higher rank is that for $\mathfrak{gl}(N)$-invariant systems there are $2^N$ independent Q-functions (see \cite{Kulish:1979cr,Kulish:1985bj,Krichever:1996qd,Dorey:2006an,Kazakov:2007fy,Gromov:2007ky}  and  \cite{Kazakov:2015efa,Gromov:2017blm,Kazakov:2018hrh} for a recent pedagogical introduction), and it not clear a priori which of them should enter the factorised expression for the Bethe states generalising \eq{eq:fact1}. 
A convenient way to label the Q-functions is by using completely antisymmetric multi-indices\footnote{At the same time there are only $N-1$ Q-functions whose roots appear in the nested Bethe ansatz equations. These are for example $Q_1,\;Q_{12},\;\dots,\;Q_{12\dots N-1}$. However, the nesting procedure contains ambiguity and can generate a number of equivalent sets of equations. Considering all such possibilities we will recover all $2^N$ Q-functions. For more details see e.g. \cite{Gromov:2017blm}.}
 \beq
Q_{i_1 \dots i_k }  \;\;,\;\; i_n \in \left\{ 1,\dots, N\right\}\;.
 \eeq
The answer to the question about which Q-functions should appear in the factorization of the Bethe states  was obtained in \cite{Gromov:2016itr} for the case of compact spin chain in the  fundamental representation of $\sl(N)$. Firstly, it was demonstrated that the Bethe states can be constructed as
 \beq\label{eq:statesIntro}
| \Psi \rangle = \prod_k\hat B^{\text{good}}(u_k) | 0 \rangle \;,
\eeq
where $\hat B^{\text{good}}(u)$ is a degree $L\times(N-1)$  polynomial in $u$ operator\footnote{When building this operator it is important to introduce an extra similarity transformation of the monodromy matrix. Such a transformation was also studied for the $\sl(2)$ case in \cite{Sklyanin:1989cg} for a slightly different model.}, such that it commutes with itself for different values of $u$. Importantly, $u_k$'s are the roots of the $Q_1$ polynomial Q-function\footnote{In fact depending on the choice of the reference state $|0\rangle$ one can use roots of any $Q_i$, the Q-function with one index.}. Following the same procedure as in \cite{Sklyanin:1991ss,Derkachov:2002tf} for $\sl(2)$ case one can label the left eigenstates of the operator $\hat B^{\rm good}(u)$ by a set of $L\times (N-1)$
real numbers ${\bf x}\equiv \{x_{i,a}\}$ with $i=1,\dots,L$ and $a=1,\dots, N-1$, such that 
\beq
\langle { \mathbf x } | \hat B^{\rm good}(u)=\prod_{i=1}^L\prod_{a=1}^{N-1}(x_{i,a}-u)\langle {\mathbf x }| \;,
\eeq
from which, together with \eq{eq:statesIntro}, it immediately follows that the Bethe state $|\Psi\rangle$ does indeed factorize into the product of $Q_1(x_{i,a})$ in this basis,
\beq\label{eq:factN}
\langle {\bf x}|\Psi\rangle = \prod_{i=1}^L\prod_{a=1}^{N-1}Q_1(x_{i,a}) \;,
\eeq
thus generalizing \eq{eq:fact1}. These results were later proven and shown to hold beyond the fundamental representation, first for $\sl(3)$ in \cite{Liashyk:2018qfc} and then for $\sl(N)$ in  \cite{Ryan:2018fyo} where the spectrum of separated variables ${\bf x}$ in more general cases was also obtained\footnote{ The form of the SoV basis $\ket{x_i}$ for the noncompact $\sl(3)$ case was elucidated recently in \cite{Derkachov:2018ewi}, while for compact spin chains an alternative construction was  proposed in  \cite{Maillet:2018bim,Maillet:2018czd,Maillet:2018rto}. Some related results for the noncompact case such as the construction of Q-operators were presented in \cite{Derkachov:2010qe,Derkachov:2006fz,Derkachov:2005hw}. }. The eigenstates construction \eq{eq:statesIntro} was extended to the supersymmetric case  in \cite{Gromov:2018cvh}.

However, the factorisation property \eq{eq:factN} does not guarantee 
the existence of a simple formula for the scalar product. The main difficulty is that $\hat B^{\rm good}$ is not self-conjugate, thus its left and right eigenvectors are not simply Hermitian conjugate to each other. This implies that the bra $\langle \Psi|$ and ket $|\Psi \rangle$ Bethe states cannot be simultaneously factorised in the same way \eq{eq:factN}.
Alternatively, one can ensure the factorization property by using the Hermitian conjugate of (\ref{eq:fact1}); however the completeness relation for $|{\bf x}\rangle$ and $(|{\bf x}\rangle)^\dagger$ is not diagonal since there is no reason to expect that $|{\bf x}\rangle$ and $(|{\bf\tilde x}\rangle)^\dagger$ are  orthogonal for ${\bf x}\neq{\bf\tilde x}$, meaning that the measure would depend on the two sets of variables $M({\bf x},{\bf \tilde x})$ giving a much more complicated expression for the norm.

In this paper, we find a different argument, independent on the explicit construction of separated variables, leading us to a concise proposal for a formula generalising (\ref{eq:mu0}) at any rank. Our derivation is based only on the Baxter TQ relations.
The main difference with the approach based on $\hat B^{\rm good}$, described above, is that our result indicates that the factorization of the bra and ket states takes place in a more intriguing way -- whereas one state still factorizes into the product of $Q_1$'s as in \eq{eq:factN}, the other state necessarily decomposes into a different set of factors. More precisely we find
\beq\label{eq:Mhat}
 {\langle }\textcolor{blue}{\Psi_A }{|}  \textcolor{cadmiumgreen}{\Psi_B } \rangle  = \int\( \prod_{a=1}^{N-1} \prod_{i=1}^{L}   d x_{i, a} \) \(  \textcolor{blue}{  \underbrace{ \prod_{a=1}^{N-1} \prod_{i=1}^{L}  Q_1^{(A)}(x_{i,a} ) }_{\text{state A} } }  \) \, \textcolor{red}{   \hat  M(  \mathbf{x} ) } \,  \(    \textcolor{cadmiumgreen}{ \underbrace{ \prod_{a=1}^{N-1}\prod_{i=1}^L Q_{\bar{a} }^{(B)}(x_{i,a} )  }_{\text{state B}} }  \) \;,
\eeq
where $Q_{\bar a}$ are the Q-functions containing the Bethe roots at the deepest level of nesting. Explicitly,
\beq\label{eq:abar}
Q_{\bar a}\equiv \frac{\epsilon^{ b_1,\dots,b_{N-1}, N+1-a}}{(N-1)!}Q_{b_1,\dots,b_{N-1}}\;
,
\eeq
and the analogue of Sklyanin's measure is  a state-independent operator acting on the wave function for one of the states. 
Like the Sklyanin's measure is a determinant of $L\times L$
matrix \eq{eq:musl2}, $\hat M({\bf x})$ is a determinant of a $L\times (N-1)$-dimensional matrix.
For instance, for $\sl(3) $ we only have  two functions $Q_{\bar 1}=Q_{12}$
and $Q_{\bar 2}=Q_{13}$ and the measure factor takes the form
\beq\la{M2}
\hat M( \mathbf{x}  ) \equiv 
\sym\det\begin{vmatrix}
 \left( \frac{x_{i,1}^{j-1} \, {\cal D}_{x_{i,1}}  }{1 + e^{2 \pi (x_{i,1} - \theta_i ) }  } \right) & \left(\frac{x_{i,1}^{j-1} \, {\cal D}_{x_{i,1}}^{-1}  }{1 + e^{2 \pi (x_{i,1} - \theta_i ) }  } \right) \\ 
 \left( \frac{x_{i,2}^{j-1} \, {\cal D}_{x_{i,2}}  }{1 + e^{2 \pi (x_{i,2} - \theta_i ) }  } \right) & \left(\frac{x_{i,2}^{j-1} \, {\cal D}_{x_{i,2}}^{-1}  }{1 + e^{2 \pi (x_{i,2} - \theta_i ) }  } \right) 
\end{vmatrix}_{1 \leq i,j\leq L} \;,
\eeq
where ${\cal D}_{x} $ is the shift operator in variable $x$:
\beq\label{eq:shiftD}
{\cal D}_x \circ f(x) \equiv f(x + i/2) \;.
\eeq
For illustration we write this result explicitly in the simplest case of $L=1$ ${\sl(3)}$ spin chain in appendix \ref{app:ex}. Schematically, we can represent (\ref{M2}) as the determinant of a tensor product
\beq
 \hat M( \mathbf{x} ) = \sym \det\begin{vmatrix} \underbrace{\left( \frac{ \hat{x}^{j-1} \,   }{1 + e^{2 \pi ( \hat x - \theta_i ) }  } \right) }_{1 \leq i,j \leq L} \otimes \left( \begin{array}{cc} { \cal D }_x & { \cal D }_x^{-1} \\
{ \cal D }_x & { \cal D }_x^{-1}\end{array} \right) \end{vmatrix} \;,
\eeq
where the first factor is the matrix appearing in the standard $\sl(2)$ Sklyanin's measure. In this form the generalization to any rank is simply
\beq
\label{measfin}
 \hat M( \mathbf{x} ) = \sym \det\begin{vmatrix} \underbrace{\left( \frac{ \hat{x}^{j-1} \,   }{1 + e^{2 \pi ( \hat x - \theta_i ) }  } \right) }_{1 \leq i,j \leq L}  \otimes \underbrace{\left( \begin{array}{cccc} { \cal D }_x^{N-2} & { \cal D }_x^{N-4} & \dots & {\cal D}_x^{2 - N} \\
 \vdots & \vdots & \ddots & \vdots\\
 { \cal D }_x^{N-2} & { \cal D }_x^{N-4} & \dots & {\cal D}_x^{2 - N}\end{array} \right)}_{\text{$(N-1)\times (N-1)$}} \end{vmatrix} \;.
\eeq
It would still be interesting to derive this formula starting from \eq{eq:statesIntro}.
It should involve the construction of a new operator, which we can tentatively denote $\hat C^{\rm good}(u)$, which would also create the states but when acting from the  right on the vacuum and evaluated at the roots of the $Q_{\bar a}$'s.
This would lead to a rigorous derivation of the scalar product we found in this paper.

 For concreteness, in this paper we exemplify the method in the case of non-compact spin chains in a specific representation\footnote{In our notation the fundamental representation has highest weight $(1,0,\dots,0).$}  with highest weight $(-1,0, \dots, 0)$. 
 We expect that the argument can be generalized to other representations, as well as to the compact case, and to be applicable also in the case of the fishnet model \cite{Gurdogan:2015csr} and $\mathcal{N}$=4 super Yang-Mills. In fact methods similar to the ones used in this paper already played a role in the computation of certain three point functions in these theories \cite{Cavaglia:2018lxi,preparation}, and we expect this extension to higher rank will help to develop  a SoV approach to the computation of correlation functions. 
 
 Let us mention that the rough structure of the general type \eq{measfin} was anticipated in \cite{SmirnovQuantM} (for a different model) based on hints from the classical SoV construction. In particular, the presence of the shift operators is nicely suggested by the classical picture (see also \cite{SmirnovClassM}).
 
 The paper is organised as follows. In section \ref{sec:strategy} we discuss in more detail our strategy and outline the derivation of our results. 
 In section \ref{sec:sl2} we discuss the simplest example of the $\sl(2)$ spin chain, for which in particular we reproduce the known Sklyanin's measure, which was obtained before in \cite{Derkachov:2001yn,Derkachov:2002wz} via a highly involved computation. Then in section \ref{sec:sl3} we derive the scalar product for the first higher rank case $\sl(3)$ and discuss in detail various complications which we will see are neatly resolved by using algebraic properties of transfer matrices. Finally in section \ref{sec:slN} we generalize our results  to any $\sl(N)$. We summarize and conclude in section \ref{sec:conclusions}.

\section{General strategy and  notation}\label{sec:strategy}
In this section we briefly outline the strategy, which we use in the rest of the paper to derive our main result -- the expression for the measure factor in the scalar product in separated variables given in \eq{eq:Mhat}. We skip most of the technical details here.

\paragraph{Q-functions and Bethe roots.}
The most familiar approach to the spin chains is in terms of the Bethe ansatz, which is a set of algebraic equations on the Bethe roots $u_{k,\alpha}$, where $\alpha=1,\dots,N-1$ represents the nesting level \cite{SutherlandN1,Kulish:1983rd}.
The lowest level roots $u_{k,1}$ are the \emph{momentum-carrying Bethe roots}: they  play a special role as the energy and momentum of a Bethe states can be expressed solely in terms of those.
Instead of using explicitly the Bethe roots it is much more convenient to pack them into the Q-functions also known as twisted Baxter polynomials. In particular $Q_1=e^{u\phi_1}\prod_k(u-u_{k,1})$, where $e^{i\phi_1}$ is an eigenvalue of the twist matrix for the quasi-periodic chain. One can show that the twisted polynomial $Q_1$ uniquely identifies the Bethe state (by twisted polynomial we mean polynomial times exponent).
Another important set of objects, which will play the central role, are the Q-functions with $N-1$ indices, which we denote by $Q_{\bar a}$ as in \eq{eq:abar}. Those contain the roots on the last level of nesting $u_{k,N-1}$, however, there is a number of ways the Bethe ansatz equations can be written, which results in $N-1$ different sets of roots at the level $\alpha=N-1$, which are labelled by the index $a=1,\dots,N-1$ of $Q_{\bar a}$. The Bethe roots $u_{k,N-1}$ {\it do not} characterise the state uniquely, as there are plenty of states with no roots at the last nesting level. However, the set of all $Q_{\bar a},\;a=1,\dots,N-1$ does determine the state uniquely as is clear from the identity 
\beq\la{Qdet}
Q_{1}=\det \left( Q_{\bar{a}}^{[ N -2 b]} \right)_{1 \leq a,b \leq N-1} \;,
\eeq
where we introduced the notation
\beq
f^{[n]} \equiv f(u + i n/2 ), \ \ f^\pm\equiv f(u\pm i/2) \;
\eeq
for the shifts in the argument. The relation \eq{Qdet} follows from the QQ-relations, see e.g. the review \cite{Gromov:2017blm}.

\paragraph{Baxter TQ relations.}
In order to find the Q-functions $Q_1(u)$, one should solve an $N$-th order finite-difference Baxter equation, which schematically has the form
\beq
    \hat O\circ Q_1=0 \;, \label{eq:Bax1}
\eeq
where the difference operator $\hat O$ is given by 
\beq\la{Of}
   \hat O\circ f\equiv 
   \tau_0^{[N]}f^{[N]}
   - 
     \tau_1^{[N-2]}f^{[N-2]}
     +\dots +  (-1)^{N-1}\tau_{N-1}^{[-N+2]}f^{[-N+2]}+
(-1)^N\tau_N^{[-N]}f^{[-N]}
\eeq
and the coefficients $\tau_k(u)$ for a spin chain of length $L$ are $L^{\rm th}$ order polynomials, which are\footnote{up to shifts of the argument and trivial overall factors} the eigenvalues of the spin chain transfer matrices in the finite-dimensional antisymmetric $\sl(N)$ irreps (see e.g. \cite{Chervov:2006xk}). The first and the last coefficients are related by 
$\tau_0(u+i/2)=\tau_N(u-i/2)$ and fixed to be the same for all states, which allows to introduce the polynomial $Q_\theta$ such that $\tau_0=Q_\theta^-, \ \tau_N=Q_\theta^+$. The other polynomials $\tau_a$, whose expansion in $u$  yields the integrals of motion for the state under consideration,
\beq\label{IMs}
\tau_a(u) = u^L \chi_{a}(G) + \sum_{j=1}^{L} u^{j-1} \, I_{a, j-1}   ,\,\,\,\, a=1,\dots, N-1 \;,
\eeq
have to be determined from the self-consistency of the equations \eq{eq:Bax1} with certain polynomiality conditions for the Q-functions.
The leading $u^L$ coefficient in $\tau_a(u)$ is  the character of the $a$-th antisymmetric $\mathfrak {su}(N)$ representation $\chi_{a}(G)$ of the diagonal $SU(N)$ twist matrix $G={\rm diag}\(
e^{i\phi_1},\dots,e^{i\phi_N}
\)$\footnote{More precisely we have $\det(1+\lambda G)=\sum_{a=0}^N\chi_a(G)\lambda^a$.}.
In the generic situation we will find only one twisted polynomial solution $Q_1(u)$ to the difference equation \eq{eq:Bax1}, which would determine us the momentum-carrying Bethe roots.
The polynomial $Q_\theta$ determines the system and its roots have the meaning of inhomogeneities. In the simplest case of the homogeneous spin chain $Q_\theta=u^L$. However, the expressions we obtain are more natural in the most general case when
\beq\label{Qtheta}
Q_\theta=\prod_{i=1}^L(u-\theta_i) \;,
\eeq
with all $\theta_i$ taken different.

It happens that the ``dual" set of 
Q-functions $Q_{\bar a}$
satisfies a very similar finite difference equation
\beq
    \hat {\widebar O}\circ Q_{\bar a}=0, \ \ \ a=1,2,\dots,N-1 \;,
\eeq
with
\beq
   \hat {\widebar O}\circ  g\equiv 
\tau_0g^{[-N]} - \tau_1g^{[-N+2]} + \dots +  (-1)^{N-1}\tau_{N-1}g^{[N-2]} +    (-1)^N\tau_N g^{[N]}\;. \label{eq:Bax2}
\eeq
Again, one can check that the equation \eq{eq:Bax2}, where the polynomial coefficients are already fixed to be the same as in \eq{eq:Bax1}, in general has $N$ independent solutions, but only $N-1$ of them can be chosen to be twisted polynomials, which are precisely our $Q_{\bar a},\;a=1,\dots,N-1$.

\paragraph{The strategy.}
By looking at \eq{eq:Bax1} and \eq{eq:Bax2} we may notice that the operators $\hat O$ and $\hat {\widebar O}$ are in a sense conjugate to each other. Indeed, they contain the same coefficients, but with different shifts in the argument. 
To make this idea more precise we have to define an inner product for  functions of one variable (which should not be confused with the scalar product of two spin chain states).
The key observation is that in fact one can define not just one but $L$ such inner products 
\beq
    \langle g f\rangle_j\equiv\int_{-\infty}^{+\infty}du\;g(u)\mu_j(u) f(u) \ , \ \ j=1,\dots,L\;.
\eeq
Below we specify more precisely the explicit form of the factors $\mu_j(u)$ defining the norm, at this stage we notice that $\mu_j(u)$ should be  $i$-periodic functions. In this case, assuming we can move the integration contour up and down in the complex plane by a multiple of $i$, we can transfer the shifts from $f$ onto $g$ while also modifying the shifts in the polynomial coefficients $\tau_n$. In this way we precisely obtain that
the two finite difference operators $\hat O$
and $\hat {\widebar O}$ are indeed conjugate to each other with respect to these inner products, \beq
\label{cong}
    \langle f\hat O g\rangle_j= 
   \langle g\hat {\widebar  O} f\rangle_j\; , \;\;\;\; 1 \leq j \leq L \;.
\eeq
In practice the exact form of the factors $\mu_j$ is constrained by the possibility to move the contours up and down and by the convergence of the integral when $f$ and $g$ are twisted polynomials. In addition, the proof of (\ref{cong}) involves certain identities, which $f$ and $g$ should satisfy, which luckily do hold in the situations where we use this argument below. In the next sections we provide more details on this and show how our approach works in explicit examples.

Having the conjugation property \eq{cong}, we can use standard arguments to prove ``orthogonality'' conditions that are satisfied by the Q-functions $Q^{A}$ and $Q^B$ corresponding to different Bethe states  $|\Psi_A \rangle $ and $|\Psi_B \rangle $. 
In fact, using (\ref{cong})  we immediately derive  $L \times (N-1)$  independent equations
\beq
\label{Odif}
    \langle Q_1^A\;(\hat {\widebar O }^{A}-\hat {\widebar O }^{B})\circ Q_{\bar a}^B\rangle_j=0\;, \;\;\;\; 1 \leq a \leq N -1\; , \;\;\;\; 1 \leq j \leq L \; ,
\eeq
where $  \hat {\widebar O }^{A} $ and $\hat {\widebar O }^{B} $ are the Baxter operators (\ref{eq:Bax2})  defined in terms of the transfer matrix eigenvalues $\tau_a^{A}(u)$ and $\tau_a^{B}(u)$  for the two states. 
Notice that \eq{Odif} can also be viewed as a linear system of equations on the coefficients of the polynomials $\tau_a^{A}(u)-\tau_a^{B}(u)$. At least one of these coefficients is nonzero whenever we consider two different Bethe states.
We have $N-1$ non-trivial polynomials  $\tau_a$ of degree $L$, which makes in total around $\sim (N-1)\times L$ non-trivial coefficients\footnote{In order to make the precise counting one needs to consider the generic {\it twisted} boundary conditions.}.
At the same time we have exactly $(N-1)\times L$ equations \eq{Odif}. In order for this homogeneous system to have a non-zero solution we must have the determinant of the system to be zero for $A\neq B$:\beq
    \det_{(a,i),(b,j)}\(\langle Q_1^A u^i Q_{\bar a}^{B [N-2 b]}\rangle_j\)  \;\propto\; \delta_{AB}\;,
\eeq
where we use the above notation to denote the determinants of the $L \times (N-1)$ matrix defined by blocks. Explicity,
\beq
\begin{vmatrix}
	\left(\<Q_1^A Q_{\bar{1}}^{B\, [N -2] } \, u^{j-1} \>_{i} \right) & 	\left(\<Q_1^A Q_{\bar{1}}^{B\, [N -4] } \, u^{j-1} \>_{i} \right)  &\dots & 	\left(\<Q_1^A Q_{\bar{1}}^{B\, [-N +2] } \, u^{j-1} \>_{i} \right) \\
		\left(\<Q_1^A Q_{\bar{2}}^{B\, [N -2] } \, u^{j-1} \>_{i} \right) & 	\left(\<Q_1^A Q_{\bar{2}}^{B\, [N -4] } \, u^{j-1} \>_{i} \right)  &\dots & 	\left(\<Q_1^A Q_{\bar{2}}^{B\, [-N +2] } \, u^{j-1} \>_{i} \right) \\
	\vdots & \vdots & \ddots & \vdots \\
	\left(\<Q_1^A Q_{\overline{N-1}}^{B\, [N -2] } \, u^{j-1} \>_{i} \right)& 	\left(\<Q_1^A Q_{\overline{N-1}}^{B\, [N -4] } \, u^{j-1} \>_{i} \right)  &\dots & 	\left(\<Q_1^A Q_{\overline{N-1}}^{B\, [-N +2] } \, u^{j-1} \>_{i} \right)
	\end{vmatrix}_{1 \leq i,j \leq L} \propto\; \delta_{AB}\;,\label{eq:Nstratgegy} 
\eeq
which leads to the Sklyanin-like scalar product defined as the rhs of (\ref{eq:Mhat}). 
Therefore, we have proved  rigorously that this expression satisfies a crucial  property for the scalar product $\langle \Psi_A | \Psi_B \rangle$:  it vanishes for any two different Bethe states. 
This derivation also reproduces the $\sl(2)$ result as its particular case $N=2$.

To offer more justification to why the proposed expression \eq{eq:Mhat}
is the scalar product in the SoV basis, we will also consider the computation of a physically well defined quantity that is easily obtainable in our formalism.
Namely, we compute the matrix element of the derivative of one of the conserved quantities $I_n$ with respect to some parameter $p$ (it could be the twist angle or the inhomogeneity). 
Whereas $\hat I_n$ itself acts diagonally on the Bethe states,
its derivative $\d \hat I_n/\d p$ is not diagonalized by the Bethe states. By computing the expectation value of this operator on a Bethe state we found that the result is a ratio of two determinants. The one in the denominator again precisely coincides with the scalar product \eq{eq:Mhat} for $A=B$.

Note that the condition of orthogonality, which our result does obey, is extremely constraining. The spectrum of the spin chain contains infinitely many states and even more distinct pairs of states, which imposes infinitely many conditions on the state-independent operator $\hat M(x)$. Given its amazingly simple form \eq{M2} and the fact that it reduces to the known norm in the $N=2$
case, there is little doubt in the validity of our result.
Nevertheless, it would be interesting to develop a rigorous derivation, which would involve explicit construction of the operator $\hat C^{\rm good}$ and its spectrum as described in the introduction.

\section{Sklyanin measure for $\sl(2)$ revisited}\label{sec:sl2}
In this section we pedagogically describe how our method works for the simplest example, namely the $\sl(2)$ noncompact rational spin chain. We will consider the case when at each site of the spin chain we have an infinite-dimensional $s=-1/2$ representation of $\sl(2)$. The Bethe ansatz equations in the most general case of inhomogeneous 
spin chain with twisted boundary conditions are
\beq
\label{baesl2}
    \prod_{n=1}^L\frac{u_j-\theta_n-i/2}{u_j-\theta_n+i/2}=-e^{2i\phi}\prod_{k=1}^M\frac{u_j-u_k+i}{u_j-u_k-i}\;,\;j=1,\dots,M\;,
\eeq
where the $\theta_n$'s are the fixed inhomogeneities at each site, which we assume to be real\footnote{this does not reduce the generality of our results as one can always analytically continue the result in $\theta_i$'s, treating carefully the integration contours.},   and $\phi$ is the fixed twist parameter. We assume  $\phi$ is nonzero and real and then as we see from (\ref{baesl2}) we can always restrict to $\phi \in (0, \pi)$. The spectrum of integrals of motion is  determined in terms of the Bethe roots $u_j$, which one can find from \eq{baesl2}.

In order to define the eigenvalues of the integrals of motion we introduce the {\it twisted} Baxter polynomial\footnote{We use the subscript $Q_1$ to emphasize the fact that the Q-system contains in this case two Q-functions. Indeed the Baxter equation has a second independent solution, $Q_2$, which contrary to $Q_1$ is not a twisted polynomial but instead has poles. }
\beq
\label{twpol}
    Q_1(u)=e^{u\phi}\prod_i(u-u_i)\;,
\eeq
then the eigenvalues of the transfer matrix can be deduced from the Baxter TQ relation
\beq
\label{Baxsl2}
\hat O\circ Q_1\equiv    Q_\theta^+ Q_1^{++} -\tau_1\; Q_1+ Q_\theta^-Q_1^{--}=0\;,
\eeq
with $Q_\theta$ defined as before in \eq{Qtheta}.
The transfer matrix eigenvalue  $\tau(u)$ is a polynomial in $u$ of the form
\beq
\label{Tsl2}
    \tau(u)=2\cos\phi\; u^L+\sum_{n=0}^{L-1}I_{n}u^{n}\;.
\eeq
Notice that its first coefficient is fixed by \eq{twpol} together with the Baxter equation. The remaining coefficients $I_n$ correspond to eigenvalues of the nontrivial integrals of motion, which in general are different for different states. 

The Baxter equation \eq{Baxsl2} is equivalent to the initial set of Bethe ansatz equations \eq{baesl2} after imposing polynomiality of $\tau_1(u)$ and also requiring $Q_1(u)$ to be of the form $e^{u\phi}\times\left[{\rm polynomial}\right]$. 
Under these conditions the Baxter equation \eq{Baxsl2} has a discrete set of solutions which are in one-to-one correspondence with the states of the spin chain. 

\subsection{Orthogonality relations}
In this section we describe yet another way of finding the
Bethe roots or equivalently $Q_1(u)$.
One notices that the Q-function has many similarities with orthogonal polynomials. 
For instance, for the case of spin chain of length $L=1$ one can show that the polynomials $q(u)=e^{-\phi u}Q_1(u)$ are  orthogonal polynomials for the measure $\frac{e^{2 u\phi}}{1+e^{2\pi (u-\theta_1)}}$. More precisely
\beq\la{ort}
\<Q_1^A\;Q_1^B\>\equiv \int_{-\infty}^{\infty}
\mu(u){Q_1^A(u)Q_1^B(u)}du\;\;\propto\;\; \delta_{AB}\;\;,\;\;
\mu(u)\equiv 
\frac{1}{1+e^{2\pi (u-\theta_1)}}
\;.
\eeq
First, we see that the integral above is convergent due to the choice $0<\phi<\pi$\footnote{The untwisting limit $\phi=0$ can be derived in a similar way.}.
Second, the orthogonality relation actually defines the polynomials $q(u)$ uniquely for a given degree of the polynomial. Thus  \eq{ort} is an alternative way of writing the Bethe ansatz equations for $Q_1(u)$.

For general $L>1$ there is more than one solution of the Bethe equations for a given number of roots, so the strict analogy with the orthogonal polynomials does not go further in the naive way.
To understand how that works, let us first derive \eq{ort} for $L=1$ from the Baxter equation, which defines for us a finite difference operator $\hat O$ \eq{Of} (which itself depends on the Bethe state through $\tau(u)$) such that $\hat O\circ Q_1=0$.
As we discussed in section~\ref{sec:strategy} there is a second operator $\hat {\widebar O}$ which in general annihilates the dual Q-functions $Q_{\bar a}$, but in the case of  $\sl(2)$ there is only one $Q_{\bar 1}$ and it coincides with $Q_1$. In other words we need to show that $\hat O$ is self-conjugate under the scalar product \eq{ort}, i.e. for any\footnote{their twists should be such that the integral is convergent} twisted polynomials $F_1,F_2$
\beq
\label{sasl2}
    \langle F_1\;\hat O\circ F_2\rangle=\langle F_2\;\hat O\circ F_1\rangle\;.
\eeq
We start from
\beqa
\label{OF2}
    \langle F_1\;\hat O \circ F_2\rangle=
    \int du\; \mu(u) F_1(Q_\theta^+F_2^{++}+Q_\theta^-F_2^{--}-\tau F_2)\;.
\eeqa
Shifting the integration contour by $-i$ in the first term (i.e. replacing there $u\to u-i$) and by $i$ in the second term, we find that this expression becomes
 precisely $\langle F_2\;\hat O \circ F_1\rangle$ as we wanted. However, we should justify the possibility to shift the integration contour. When doing the shift $u\to u-i$ for the first term in r.h.s. of \eq{OF2}, we should be careful as we are moving the contour through the point $u=\theta_{1}-i/2$ where the measure $\mu$ has a simple pole. However, we have chosen this pole to be precisely at the location where the factor $Q_\theta^+$, originating from the Baxter equation, has a zero. Thus we can indeed move the contour. The same argument applies to the second term in \eq{OF2}, in which the pole at $u=\theta_{1}+i/2$ is compensated by $Q_\theta^-$. As a result, \eq{sasl2} is indeed valid. Now, the proof of the orthogonality \eq{ort}
 is almost immediate:
\beqa\la{derL1}
0&=&0-0=\<Q_1^A\; \hat O^B \circ Q_1^B\>-\<Q_1^B \;\hat O^A \circ Q_1^A\>=
\<Q_1^A \;(\hat O^B-\hat O^A) \circ Q_1^B\>\\
\nn &=&
\<Q_1^A (\tau^A-\tau^B) Q_1^B\>
=
(I^A_0-I^B_0)\<Q_1^A  Q_1^B\>\;,
\eeqa
where we added the superscripts $A$ and $B$ to indicate that $\hat O$ is different for the two different states. Finally, we note that for two different states the values of integrals of motion  $I_0$ have to be distinct, leading to the conclusion that \eq{ort} is indeed true.

\paragraph{Orthogonality for general $L$.}
Now we can see the difficulty one would have for $L>1$. The self-conjugation property of $\hat O$ would be still valid and all steps in \eq{derL1} would go through, except for the last one.
What we get instead is
\beq\la{ort2}
\<Q_1^A (\tau^A-\tau^B) Q_1^B\>
=
\sum_{i=0}^{L-1} (I^A_i-I^B_i)\<Q_1^A u^i Q_1^B\>=0\;,
\eeq
which no longer implies \eq{ort}.
However, for $L>1$ we also gain a freedom in how to define $\mu(u)$. Namely, we can use any of the following measures:
\beq
\mu_j(u)=\frac{1}{1+e^{2\pi (u-\theta_j)}}\;\;,\;\;j=1,\dots, L\;,
\eeq
 and we will denote the corresponding integrals as
\beq\label{eq:bracj}
\<f\>_j\equiv \int_{-\infty}^\infty \mu_j(u) f(u) du \;.
\eeq
This means that we have a set of $L$ equations like \eq{ort2},
\beq\label{ort3}
\sum_{i=1}^{L} (I^A_{i-1}-I^B_{i-1})\<Q_1^A\; u^{i-1} Q_1^B\>_j=0\;\;,\;\;j=1,\dots,L\;, 
\eeq
where $\<\cdot\>_j$ is defined by \eq{eq:bracj}. 
This homogeneous system of equations is only compatible if the determinant of the linear system is zero, so we get
\beq\label{eq:matdet}
\text{det} \left|\<Q_1^A\; u^{i-1} Q_1^B\>_j \right|_{1\leq i,j \leq L} \;\propto\;\delta_{AB}\;.
\eeq
Note that each entry in the matrix \eq{eq:matdet} is defined as a single integral. However, we can rewrite it in the form that leads precisely to Sklyanin's scalar product for $\sl(2)$,
\beq\label{eq:Sklya1}
\int \prod_{i=1}^L {d x_i} \;
Q^A_1(x_i) M({\bf x}) Q^B_1(x_i) \;\propto \;\delta_{AB}\;,
\eeq
where we also use that we can symmetrize over  the integration coordinates $x_i$,
ensuring that the measure factor $M$ is symmetric in its $L$ arguments:
\beq\label{eq:muL}
M({\bf x}) =\frac{ \prod\limits_{ j<k} (e^{2\pi x_j}-e^{2\pi x_k})(x_j-x_k)}{ \prod\limits_{j,k}  (1 + e^{2\pi (x_j-\theta_k) } )}\;,
\eeq
which is precisely the measure derived in \cite{Derkachov:2001yn,Derkachov:2002wz}.  
So, in conclusion, we have re-derived the orthogonality
of the Bethe states $|\Psi\>$ written in the SoV basis via \eq{ort3}.
We can now change the direction of the logic and declare that the orthogonality relation \eq{eq:Sklya1} is a way  alternative to \eq{baesl2} for defining this system from which one can determine $Q_1(u)$ and thus find the spectrum. We see that the knowledge of the measure is a powerful and non-trivial seed containing the knowledge of the spectrum. In addition one can utilize it to compute some non-trivial matrix elements as we show in the next section.

\subsection{Simple form factors}\label{sec:FFs}
Some form-factors, such as 1-point functions in 2D Sinh-Gordon theory,
can be expressed in a nice way in terms of the Sklyanin's type of measure~\cite{Lukyanov:2000jp}. Generalization of this approach could lead to a non-perturbative expression of some $3$-point functions in much more complicated theories such as 4D ${\cal N}=4$ SYM.
Here we consider a prototype of such observable -- a diagonal 
matrix element of the variation of an integral of motion $\hat I_n$ w.r.t. a parameter $p$. In $\mathcal{N}$=4 SYM for instance one could consider the variation  of the dilatation operator with respect to the coupling constant. The corresponding expectation values are associated to diagonal OPE coefficients involving the Lagrangian $\mathcal{L}$\footnote{More generally, in any CFT one can obtain in this way diagonal OPE coefficients $\mathcal{C }_{\mathcal{O}\mathcal{O} \mathcal{M}}$, involving a generic operator $\mathcal{O}$ and a marginal operator $\mathcal{M}$~\cite{Costa:2010rz}.}. In some limits of this theory studied together with A.~Sever we indeed found a formula reminiscent of SoV for this observable \cite{Cavaglia:2018lxi, preparation}. Here we generalise the method introduced there.   

For simplicity in the present setting we consider variations with respect to the twist, $p=\phi$. To study the matrix element one can use the standard logic of the perturbation theory,\footnote{For real inhomogeneities and twists the coefficients $\hat I_n$ in the transfer matrix should be linear combinations of
mutially commuting self-conjugate operators.}
\beq\la{dI}
0=\d_p \<\Psi|(\hat I_n-I_n)|\Psi\>=
 \<\Psi|(\d_p\hat I_n-\d_p I_n)|\Psi\>
\eeq
meaning that the expectation value of the non-diagonal operator $\d_p\hat I_n$ is given by the derivative of the eigenvalue  $I_n$ w.r.t. the parameter, 
\beq\label{dI2}
   \frac{\langle\Psi|\frac{\d\hat I_n}{\d p}|\Psi\rangle}{\langle\Psi|\Psi\rangle}=\frac{\d  I_n }{\d p}\;.
\eeq
The r.h.s. is already much easier to compute -- one could find 
a solution of the Bethe ansatz equations at two close values of the parameter $p$ and then find the difference of the T-function coefficient. However, if we think about the l.h.s. of  \eq{dI2}
as a matrix element we should be able to write it in the SoV basis, which should look similar to the expression for the norm~\eq{eq:Sklya1} with  possible extra insertions, meaning that we should be able to express the result in terms of $Q_1$ computed at one given value of $p$.

To achieve this we can use some tricks from the previous section. Namely, consider
\beq
\label{var1}
    0=\langle Q_1\;(\hat O+\delta\hat O)\circ (Q_1+\delta Q_1)\rangle_i=\langle Q_1\;\hat O\circ \delta Q_1\rangle_i + \langle Q_1\delta\hat O \circ Q_1\rangle_i\;,
\eeq
where $\delta$ stands for the variation w.r.t. the parameter $p$.
Note that the first term vanishes since we can act with $\hat O$ to the left, as a result we get
\beq
\label{qdo}
    \langle Q_1\;(\d_p\hat O) \circ Q_1\rangle_i=0\; .
\eeq
For definiteness let us take $p=\phi$. In this case $\d_p \hat O=\d_p \tau(u)=\sum_{n=0}^{L} u^n \d_p I_{n}$.
The main difference with the previous section is that the leading term in $\tau(u)$ does not cancel, since $I_L=2\cos\phi$. This means that the system of equations \eq{qdo} is a non-homogeneous system of the form
\beq\label{eq:lsys}
	\sum_{n=0}^{L-1}  	\langle Q_1^2 \, u^{n}\rangle_i  \;\d_\phi I_{n}=
	2\sin\phi\;\langle Q_1^2 \, u^{L}\rangle_i  \ \ 
	, \ \ \ \ i=1,\dots,L\;,
\eeq
We see that the matrix in this linear system is exactly the same as in \eq{ort3} with $A=B$. 
This means that by solving the linear equation (\ref{eq:lsys}) by Cramer's rule, we  obtain an expression of the form factor in terms of a ratio of determinants, where in the denominator we have the same determinant (\ref{eq:matdet}) defining the ``square norm" of the state, and in the numerator the determinant of the same matrix, but with one column replaced:
\beq
	 \d_\phi I_k=\frac{1}{2\sin\phi}\frac{\det\limits_{i,j=1,\dots,L}m_{ij}^{(k)}}{\det\limits_{i,j=1,\dots,L}m_{ij}}
	, \;\;\; k=0,\dots, L-1\;,
\eeq
where
\beq
m_{ij} \equiv \langle Q_1^2 \, u^{j-1}\rangle_i\;\;;\;\;
m_{ij}^{(k)}=m_{ij}, \ \ \text{for}\ \ j\neq k+1\;\;\text{and}\;\; m_{i,k+1}^{(k)}= \langle Q_1^2 \, u^{ L}\rangle_i\;.
\eeq
Evaluating these determinants explicitly we get the SoV-type formula
\beq\la{dIr}
\frac{\langle \Psi| \frac{\d \hat{ I}_l }{\d \phi} | \Psi \rangle }{\langle \Psi| \Psi \rangle} =  \frac{(-1)^l}{2 \sin\phi}\;\frac{ \, \int d^L {\mathbf x} \; \Psi(\mathbf{x} ) \,  M(\mathbf{x} ) P_{L-l + 1}(\mathbf{x} ) \, \Psi(\mathbf{x} )}{ \int  d^L {\mathbf x}  \;\Psi(\mathbf{x} )\,  M(\mathbf{x} ) \, \Psi(\mathbf{x} ) }\,,
\eeq
where the wave function in separated variables is given by the factorized product of the Q-function \eq{eq:fact1}, and $P_n$ is a homogeneous polynomial of degree $n$, obtained as a symmetrized product of $n$ distinct variables from $x_1 , \dots, x_L$, with unit normalization for each monomial\footnote{e.g., for $L=3$, $P_1(\mathbf{x}) = x_1 + x_2 + x_3$, $P_2(\mathbf{x} ) = x_1 x_2 + x_1 x_3 + x_2 x_3$,  $P_3(\mathbf{x}) = x_1 x_2 x_3 $.}.

Note that in our case the insertion resulting from $\d_\phi \hat I_n$ is just a function of $\bf x$, however, it is clear that
already for $\d_{\theta_j} \hat I_n$ we would get also some shift operators
acting on one of the $\Psi(\bf x)$ in the numerator of \eq{dIr}. This is in fact a generic feature of the SoV type of integrals as we will see in the next section.

\section{SoV scalar product in $\sl(3)$ spin chains}\label{sec:sl3}

In this section we exemplify our approach for the $\sl(3)$ case. 
Our starting point is a set of nested Bethe ansatz equations~\cite{SutherlandN1,Kulish:1983rd},
\beqa\la{Bethe1}
    \prod_{n=1}^L\frac{u_j-\theta_n-i/2}{u_j-\theta_n+i/2}&=&e^{i(\phi_1-\phi_2)}\prod_{k\neq j}^{N_u}\frac{u_j-u_k+i}{u_j-u_k-i}
    \prod_{l=1}^{N_v}\frac{u_j-v_l-i/2}{u_j-v_l+i/2}\ , \\ \la{Bethe2}
    1&=&e^{i (\phi_2 - \phi_3)  }\prod_{k\neq j}^{N_v}\frac{v_j-v_k+i}{v_j-v_k-i}\prod_{l=1}^{N_u}\frac{v_j-u_l-i/2}{v_j-u_l+i/2}\;,
\eeqa
where $u_j$ are the momentum-carrying roots and $v_j$ are the auxiliary Bethe roots. We consider the quasi-periodic boundary conditions parametrized by three twist angles $\phi_i$, with $\sum_{i=1}^3 \phi_i = 0$. 

As we already mentioned in the introduction the nested Bethe ansatz is ambiguous and in the current case has an alternative ``dual" form (see e.g. \cite{Bazhanov:1996dr,Pronko:1999gh,Gromov:2007ky})
\beqa
    \prod_{n=1}^L\frac{u_j-\theta_n-i/2}{u_j-\theta_n+i/2}&=&e^{i(\phi_1-\phi_3)}\prod_{k\neq j}^{N_u}\frac{u_j-u_k+i}{u_j-u_k-i}
    \prod_{l=1}^{N_w}\frac{u_j-w_l-i/2}{u_j-w_l+i/2}\ , \\ 
    1&=&e^{i (\phi_3 - \phi_2) }\prod_{k\neq j}^{N_w}\frac{w_j-w_k+i}{w_j-w_k-i}\prod_{l=1}^{N_u}\frac{w_j-u_l-i/2}{w_j-u_l+i/2}\;,
\eeqa
where $N_w=N_u-N_v$.

As in the previous section we introduce the Baxter (twisted) polynomials
\beq
\label{Qsl3}
    Q_{1}=e^{\phi_1u}\prod_{j=1}^{N_u}(u-u_j)\;\;,\;\;  Q_{12}=e^{(\phi_1+\phi_2)u}\prod_{j=1}^{N_v}(u-v_j)
    \;\;,\;\;  Q_{13}=e^{(\phi_1+\phi_3)u}\prod_{j=1}^{N_w}(u-w_j)\;. 
\eeq
The ``dual" roots $w_k$ are not independent and can be derived from given $u_j$ and $v_k$ via the QQ-relation
\beq\la{Q1}
Q_1\;\propto\; Q_{12}^+Q_{13}^--Q_{12}^-Q_{13}^+\;,
\eeq
which is valid up to a trivial proportionality factor. 

Like in the previous section, we need to show that the Q-functions satisfy a
finite difference equation with some polynomial coefficients as we outlined in \eq{Of} and \eq{eq:Bax2}. Let us define two polynomials $\tau_1$ and $\tau_2$ \cite{Krichever:1996qd,Gromov:2010kf}
\beqa
\label{Tsl3}
    \tau_1&=& 
    Q_\theta^+\frac{Q_1^{++}}{Q_1}+ Q_\theta^-\frac{Q_1^{--}}{Q_1}\frac{Q_{12}^+}{Q_{12}^-}+Q_\theta^-\frac{Q_{12}^{[-3]}}{Q_{12}^-}    \;,\\
    \nn
    \tau_2&=&Q_\theta^+\frac{Q_{12}^{[+3]}}{Q_{12}^+}
    +Q_\theta^+\frac{Q_1^{++}}{Q_1}\frac{Q_{12}^{-}}{Q_{12}^+}
    +Q_\theta^-\frac{Q_1^{--}}{Q_1}\;.
\eeqa
One can check that these combinations of the Q-functions are indeed polynomials by observing that all poles must cancel due to the Bethe equations \eq{Bethe1} and \eq{Bethe2}.
Also, it is easy to check from \eq{Q1} that in \eq{Tsl3} one can replace $Q_{12}$ by $Q_{13}$ without changing the l.h.s.
Finally, one can see that $\tau_2$ and $\tau_1$ are complex conjugate to each other.

Having $\tau_1$ and $\tau_2$ defined, we can easily verify that
\beqa
\label{Bax1sl3}
\hat O\circ Q_1&\equiv&	Q_\theta^{++}Q_1^{[+3]} - \tau_1^+Q_1^+ + \tau_2^-Q_1^- - Q_\theta^{--}Q_1^{[-3]}=0\;,\\
\label{Bax2sl3}
\hat {\widebar O}\circ Q_{\bar a}&\equiv&     Q_\theta^-Q_{\bar{a}}^{[-3]}- \tau_1Q_{\bar{a}}^- +\tau_2Q_{\bar{a}}^+ -Q_\theta^+Q_{\bar{a}}^{[+3]} =0\; , 
\eeqa
where the second equation is satisfied by both $Q_{\bar 1}\equiv Q_{12}$ and $Q_{\bar 2}\equiv Q_{13}$.
To check \eq{Bax1sl3} and \eq{Bax2sl3} one should simply plug the definition \eq{Tsl3} into the above equations and check that all terms cancel.

As was outlined in the section~\ref{sec:strategy}, we need to demonstrate that these two finite difference operators \eq{Bax1sl3} and \eq{Bax2sl3}
are conjugate w.r.t. some inner product. Since \eq{Bax1sl3} and \eq{Bax2sl3} do have the correct form as in \eq{Of} and \eq{eq:Bax2}, this property is almost guaranteed if we are allowed to move the integration contour.
In the next section we verify that all extra contributions arising from the shifts of the contours do cancel.

\subsection{Poles cancellation}\label{sec:bracketSL3}
Like in the previous section we define the bracket
\beq
\<f\>_j\equiv \int_{-\infty}^\infty \mu_j(u) f(u) du\;\;,\;\;
\mu_j(u)=\frac{1}{1+e^{2\pi(u-\theta_j)}}\;.
\eeq
What we are going to show is that
\beq\label{weak3}
\<Q_1\;\hat {\widebar O} \circ f\>_j = 0\;,
\eeq
where $f$ is a twisted polynomial with the same asymptotic as any of $Q_{\bar a}$,
it other words we do not require the roots of $f$ to satisfy the Bethe ansatz equations,
otherwise the statement \eq{weak3} would be trivial.
First let us comment on the convergence of this integral. Assuming $f = e^{\alpha u} \times\left[\text{polynomial}\right]$, the integrand in (\ref{weak3})  goes like $ e^{(\phi_1 + \alpha - 2 \pi ) u } \times\left[ \text{polynomial}\right]$ at $u \rightarrow + \infty$ and $ e^{(\phi_1 + \alpha  ) u } \times\left[ \text{polynomial}\right]$ at $u \rightarrow - \infty$. From this we deduce the condition of convergence $0 < \phi_1 + \alpha <  2 \pi $. Since this inequality should hold for $\alpha = \phi_1 + \phi_2$ or $\alpha = \phi_1 + \phi_3$, which are the twists in $Q_{\bar{1}}$ and $Q_{\bar{2}}$, we get
\beq\la{convergence}
0 < \phi_1 - \phi_2 < 2 \pi\;\; ,\;\; 0 < \phi_1 - \phi_3 < 2 \pi\;.
\eeq
Note, that the condition \eq{convergence} does not restrict the generality of our consideration as one can always choose $\phi_a$'s so that \eq{convergence} is satisfied. The only physically distinguished combinations of the $\phi_a$'s are the phases $e^{i(\phi_1-\phi_2)}$ and $e^{i(\phi_2-\phi_3)}$, appearing in the Bethe ansatz equations \eq{Bethe1} (where we still assume that $\phi_1+\phi_2+\phi_3=0$).

To prove \eq{weak3}, we show that we can transfer $\hat {\widebar  O }$ to become $\hat O $ acting on $Q_1$, which gives zero, 
\beqa
\label{weakd}
    \langle Q_1\;\hat {\widebar O} \circ f\rangle_j&=&
\int_{-\infty}^{+\infty} \mu_j(u) Q_1(u)\[ Q_\theta^-f^{[-3]} - \tau_1f^- + \tau_2f^+ -Q_\theta^+f^{[+3]} \]du \\ \nn
    &=&\int_{-\infty+i0}^{+\infty+i0} \mu_j(u+\tfrac i2) \[\underbrace{Q_\theta^{++}Q_{1}^{[+3]} - \tau_1^+Q_{1}^+ + \tau_2^-Q_{1}^- -Q_\theta^{--}Q_{1}^{[-3]} }_{\hat O \circ Q_1 = 0 }\]f(u)\;du
    \\ \nn
    &+& {\text{residues from poles}}\;,
\eeqa
where we shifted the integration contour in each term 
so that at the end $f$ appears with no shift.
This results in shifts in $Q_1$ and we see that we get precisely the Baxter equation for $Q_1$ in the square brackets and also a shift of the argument in the $i$-periodic measure factor $\mu_j(u+\tfrac  i2)$. As a result, the only potentially nonzero contribution comes from residues at poles of the measure $\mu_j$ that we cross when shifting the contour.
The measure $\mu_j(u)$ has poles at $u=\theta_j+\tfrac i2+i n,\;n\in {\mathbb Z}$ with the same residue $-\frac{1}{2\pi}$. We are going to compute these residues.
\paragraph{Residues from the first term.} For the first term $\mu_j(u) Q_1(u) Q_\theta(u-\tfrac i2) f(u-\tfrac {3i}2)$ we will need to shift the contour up by $\tfrac{3i}{2}+0i$, so that the final integration in the second line of \eq{weakd} is slightly above the real axis.
While shifting the contour we have two potential locations of residues which can contribute to the result -- these are at $u=\theta_j+\tfrac i2$ and at $u=\theta_j+\tfrac{3 i}{2}$.
However, the first one does not contribute since $Q_\theta(\theta_j)=0$. So we are left with the contribution
\beq
r_1= -i Q_1(\theta_j+\tfrac{3i}{2}) Q_\theta(\theta_j+\tfrac{i}{2}) f(\theta_j)\;. 
\eeq
\paragraph{Residues from the second term.} For the second term $-\mu_j(u)Q_1(u)\tau_1(u)f(u-\tfrac{i}{2})$ we only have one contribution at $u=\theta_j+\tfrac{i}{2}$, which gives
\beq
r_2=+iQ_1(\theta_j+\tfrac{i}{2})\tau_1(\theta_j+\tfrac{i}{2})f(\theta_j)\;.
\eeq
Similarly, one can see there are no extra contributions from the remaining two terms
and 
we get the following result
\beq
\langle Q_1\;\hat {\widebar O} \circ f\rangle_j=r_1+r_2=
if(\theta_j)\[
Q_1(\theta_j+\tfrac{i}{2})\tau_1(\theta_j+\tfrac{i}{2})
-Q_1(\theta_j+\tfrac{3i}{2}) Q_\theta(\theta_j+\tfrac{i}{2})
\]\;.
\eeq
Finally, by looking at the definition \eq{Tsl3} of the transfer matrix eigenvalue we see that the expression in the square brackets is precisely zero. This leads to the result \eq{weak3}.

We view this cancellation of residues as a significant indication of the validity of our approach.
We see that even though those extra poles can spoil the generalisation from $\sl(2)$
to $\sl(3)$, luckily there exist these extra relations between Q-functions and transfer matrix eigenvalues, enabling the formalism to work. 

\subsection{Orthogonality relations}
As we have already explained in section \ref{sec:strategy},  the relation of the type \eq{weak3} is  the starting point for the derivation of the  scalar product.
Here we demonstrate that the general argument for the scalar product and the orthogonality relation goes through in the $\sl(3)$ case.

First, consider the relation
\beq\la{AB30}
\left\langle Q_1^A\; ( \hat {\widebar O }^A - \hat {\widebar O}^B )\circ Q_{\bar{a}}^B \right\rangle_i = 0\;\;,\;\;a=1,2\;\;,\;\;i=1,\dots,L\;,
\eeq
where we again use the superscript $A$ and $B$ to indicate that those Q-functions and Baxter operators correspond to two different Bethe states $|\Psi^A\>$ and $|\Psi^B\>$.
To prove the relation above  we use that $\hat{\widebar O}^B Q_{\bar a}^B=0$ and the property
\eq{weak3} with $f=Q_{\bar a}^B$.

Next, we use again that the first and the last terms in $\hat {\widebar O}$ do not depend on the state by definition \eq{Bax2sl3} and we get
\beq\la{AB3}
\(\hat {\widebar O}^A-\hat {\widebar O}^B\)\circ Q_{\bar{a}}^B=
\sum_{j=1}^{L}
\[- (I_{1,j-1}^A-I_{1,j-1}^B)u^{j-1}\;{\cal D}^{-1}\circ Q_{\bar{a}}^B +(I_{2,j-1}^A-I_{2,j-1}^B)u^{j-1}\;{\cal D}\circ Q_{\bar{a}}^B\]\;,
\eeq
with the shift operator defined as in (\ref{eq:shiftD}). 
Plugging \eq{AB3} into the relation \eq{AB30} we get a linear system of 
equations:
\beq\la{AB31}
\sum_{(b,j)=(1,1)}^{(2,L)}\left\langle Q_1^A \;u^{j-1}\; {\cal D}^{-3+2b}\circ Q_{\bar{a}}^B \right\rangle_i \;
\times\;(-1)^b\(I_{b,j-1}^A-I_{b,j-1}^B\)
= 0\; ,
\eeq
where we introduce the multi-index $(b,j)$, which takes $2L$ different values.
This equation tell that $(-1)^b\(I_{b,j-1}^A-I_{b,j-1}^B\)$ should be a null vector of the $2L\times 2L$ matrix. In other words the determinant of this matrix should be zero for $|\Psi^A\>\neq |\Psi^B\>$, as for two different states at least some conserved charges should be different, so we get
\beq\label{detsl3}
\det_{(a,i),(b,j)}\left\langle Q_1^A \;u^{j-1}\; {\cal D}^{-3+2b}\circ Q_{\bar{a}}^B \right\rangle_i=0\;.
\eeq
This is our orthogonality relation \eq{eq:Nstratgegy}. 
We emphasise again that the existence of a simple orthogonality 
relation is highly nontrivial as there are infinitely many states in this model.
Such an orthogonality relation should have an explanation at the level of the operators
acting on the spin chain states such as $\hat B^{\rm good}(u)$ and $\hat C^{\rm good}(u)$,  discussed in the introduction.

\section{Extension to any $\sl(N)$}\label{sec:slN}
In this section we extend the observations made in the previous section to the case of  $\sl(N)$ and prove the general
formula for the scalar product \eq{eq:Nstratgegy}. 

There are two main relations to prove. First, we have to show that the Baxter equations for $Q_1$ and $Q_{\bar a}$ are indeed of the form \eq{Of}, \eq{eq:Bax2}. Second, 
we need to demonstrate the cancellation of poles in the identity \eq{weak3} for the case of any $\sl(N)$.
After that we can can use $\eq{weak3}$ to derive the orthogonality relation between two different states in the SoV basis and read off the SoV measure from that as was done in the previous section.

\subsection{Baxter operators for $\sl(N)$ spin chain}
Here we use the general formalism which allows one to build the eigenvalues of the transfer matrices in finite-dimensional totally antisymmetric representations $\tau_k$\footnote{The actual eigenvalues $T_k$ of the transfer matrices are related to $\tau_k$
in the following way: $T_k=\prod_{l=2}^k Q^{[2l-3]}_\theta \tau_k$ and $T_1=\tau_1,\;T_0=1$.},
corresponding to Young diagrams with $k$ boxes developed in \cite{Krichever:1996qd}
(for a review see \cite{Gromov:2010kf}). In this method they are obtained from the 
generating functional
\beq\label{eq:WT}
    {\cal W}=\sum_{k=0}^N (-1)^k\tau_{k} \, {\cal D}^{2 k}\; ,
\eeq
which can be written in analogy with the generating function for characters of  antisymmetric $\sl(N)$ representations as
\beq
\label{Wsl4}
    {\cal W}=Q_\theta^-(1-R_1 {\cal D}^2)(1-R_2 {\cal D}^2)\dots (1-R_N {\cal D}^2) \;,
\eeq
where each of the factors contains the shift operator ${\cal D}$ 
and a rational function $R_i$, which is a combination of the twisted Baxter polynomials,
\beq
R_1=\frac{Q^{+}_{\theta}}{
Q_{\theta}^{-}
}
\frac{
Q_{1}^{++}
}{Q_{1}}
\;\;,\;\;R_i=\frac{
Q_{J_{i-1}}^{[-i]}
}{Q^{[2-i]}_{J_{i-1}}}
\frac{
Q_{J_{i}}^{[3-i]}
}{Q^{[1-i]}_{J_{i}}} \ \ , \ \ i=2,\dots,N \;,
\eeq
where we define the multi-index $J_i\equiv 12\dots i$, such that for example $Q_{\bar 1}=Q_{J_{N-1}}$. We also define
\beq
Q_{J_{0}}\equiv \frac{1}{Q_\theta}\;\;,\;\;Q_{J_{N}}\equiv 1\;.
\eeq
We assume that the twisted Baxter polynomials have the following form
\beq
Q_{i_1\dots i_l}=e^{u\sum_{p=1}^l \phi_{i_p}}\times \[\text{polynomial}\]\;,
\eeq
with $\sum_{a=1}^N \phi_a=0$.

We also have to show that $\tau_k$'s are actually polynomials.
This is not totally trivial as $R_i$ are rational functions with various poles. 
We need to show that these poles cancel as a consequence of the Bethe ansatz equations. Let's look at the poles related to the Bethe roots at nesting level $k$, i.e. coming from zeros of $Q_{J_k}$. There are two $R$'s which contain $Q_{J_k}$ in the denominator: $R_k$ and $R_{k+1}$. Let us focus on the two terms containing these $R$'s,
\beqa
&&\dots (1-R_k {\cal D}^2)
 (1-R_{k+1} {\cal D}^2)\dots = 
 \dots \(1-(R_k+R_{k+1}) {\cal D}^2
 +R_k R_{k+1}^{++} {\cal D}^4
 \)\dots\\
&=&
\dots \(1-\left[\frac{
Q_{J_{k-1}}^{[-k]}
}{Q^{[2-k]}_{J_{k-1}}}
\frac{
\textcolor{blue}{Q_{J_{k}}^{[3-k]} }
}{ \textcolor{blue}{Q^{[1-k]}_{J_{k}}} }+
\frac{ \textcolor{blue}{Q_{J_{k}}^{[-k-1]} }
}{ \textcolor{blue}{Q^{[1-k]}_{J_{k}}  } }
\frac{
Q_{J_{k+1}}^{[2-k]}
}{Q^{[-k]}_{J_{k+1}}}
\right] {\cal D}^2
 +\frac{
Q_{J_{k-1}}^{[-k]}
}{Q^{[2-k]}_{J_{k-1}}}
\frac{
Q_{J_{k+1}}^{[4-k]}
}{Q^{[2-k]}_{J_{k+1}}} {\cal D}^4
 \)\dots
\eeqa
We see that the poles due to zeros of $Q_{J_k}$ in the square bracket cancel if we impose at the roots of $Q_{J_k}(u)=0$
the following condition
\beq
\frac{
Q_{J_{k-1}}^{-} 
}{Q^{+}_{J_{k-1}}}
\frac{
Q_{J_{k}}^{++} 
}{Q_{J_{k}}^{--} }
\frac
{Q_{J_{k+1}}^{-}
}{Q^{+}_{J_{k+1}}}
=-1\;,
\eeq
which is exactly the Bethe ansatz equation at the nesting level $k$. This argument applies for all $k=1,\dots,N-1$. In addition we should check that the poles at $u=\theta_j+\tfrac{i}{2}$ in $R_1$ cancel, however, this pole is nicely cancelled by the $Q_\theta^-$ prefactor in ${\cal W}$.
Thus indeed all $\tau_k$'s are polynomials due to the Bethe equations, just like in the $\sl(3)$ case.

Now let us show that
\beq
\label{ObW}
    \hat{\widebar O}={\cal W}{\cal D}^{-N} \;,
\eeq
indeed we see that it annihilates $Q_{\bar 1}$,
\beqa\la{ObarW}
{\cal W}{\cal D}^{-N} Q_{\bar 1}=
{\cal W}{\cal D}^{-N} Q_{J_{N-1}}&=&
\dots (1-R_{N}{\cal D}^2)Q_{J_{N-1}}^{[-N]}\\
\nn&=&
\dots \(Q_{J_{N-1}}^{[-N]}-
\frac{
Q_{J_{N-1}}^{[-N]}
}{Q^{[2-N]}_{J_{N-1}}}
Q_{J_{N-1}}^{[2-N]}\)=0\;.
\eeqa
Furthermore, it is obvious that $\tau_0=Q_\theta^-$ and 
\beq
\tau_N=Q^-_\theta R_1 R_2^{++}\dots R_N^{[2N]}=Q^+_\theta\;.
\eeq
So indeed the Baxter equation for $Q_{\bar 1}$ is of the general form given in \eq{eq:Bax2}.
We should also show that $\hat{\widebar O}$ annihilates {\it any }   
$Q_{\bar a},\;a=1,\dots,N-1$. The remaining $Q_{\bar a}$'s are defined through the bosonic duality transformation \cite{Bazhanov:1996dr,Pronko:1999gh,Gromov:2010kf}. Like in the $\sl(3)$ case, one can show that the polynomials $\tau_k$'s are invariant under this transformation~\cite{Gromov:2010kf}. For example, the duality 
transformation which defines $Q_{\bar 2}=Q_{1,2,\dots ,N-2,N}$ is
\beq
Q_{J_{N-2}}\;\propto\; Q^+_{\bar 1}Q^-_{\bar 2}-Q^+_{\bar 1}Q^-_{\bar 2} \;,
\eeq
which leads to the following identity
\beq
(1-R_{N-1} {\cal D}^2)(1-R_{N} {\cal D}^2)=
(1-\tilde R_{N-1} {\cal D}^2)(1-\tilde R_{N} {\cal D}^2)\;,
\eeq
where  $\tilde R_i$ are the same as $R_i$ with $Q_{J_{N-1}}=Q_{\bar 1}$ replaced by $Q_{\bar 2}$. After that one can repeat the same argument as in \eq{ObarW} to show that ${\cal W}{\cal D}^{-N}Q_{\bar 2}=0$.
To obtain all $Q_{\bar a}$ one should apply the bosonic duality to other factors in $\cal W$ as well, as explained in detail in \cite{Gromov:2010kf}.

In a similar way we can construct the Baxter equation for $Q_1$. For that consider
\beq
\label{Wsl41}
    {\cal W}^\dagger\equiv (1-R^{--}_N {\cal D}^{-2})(1-R^{--}_{N-1} {\cal D}^{-2})\dots 
    (1-R_1^{--} {\cal D}^{-2})
    Q_\theta^-\;,
\eeq
which is related to ${\cal W}$ by a formal conjugation, which flips the order of 
the operators and replaces ${\cal D}$ by its inverse i.e. 
these two generating functionals are related according to the rules ${\cal D}^\dagger \equiv {\cal D}^{-1}$ and $f(u)^\dagger\equiv f(u)$ and $(A B)^\dagger\equiv B^\dagger A^\dagger$\ \footnote{This transformation is consistent with the main algebraic identity for the shifts operators ${\cal D}f=f^{+}{\cal D}$, which transforms under $\dagger$ to $f{\cal D}^{-1}={\cal D}^{-1}f^{+}$ which is equivalent to the initial one.}. Applying this operation to the 
representation of ${\cal W}$ \eq{eq:WT} we get
\beq\la{Wdag}
{\cal W}^\dagger = 
\sum_{k=0}^N (-1)^k \tau_{k}^{[-2k]}{\cal D}^{-2 k}\; .
\eeq
We now can see that ${\hat O}={\cal W}^\dagger$. Indeed
\beq
{\cal W}^\dagger Q_1
=\dots (1-R_1^{--} {\cal D}^{-2}) Q_\theta^- Q_1
=\dots \(Q_\theta^- Q_1-\frac{Q^{-}_{\theta}}
{
Q_{\theta}^{[-3]}
}
\frac{
 Q_{1}^{}
}{Q^{--}_{1}} Q_\theta^{[-3]} Q_{1}^{--}\) =0\;.
\eeq
Also we see that $\hat O$ defined in this way indeed agrees with \eq{Of} in section \ref{sec:strategy} due to \eq{Wdag}.

\subsection{Poles cancellation}
We have to demonstrate that the relation \eq{weak3} still holds for general $\sl(N)$. First, we need to ensure the convergence of the integral \eq{weak3}. This time we assume that $f(u)$ can be of the form $e^{-u\phi_{c}}\times[\text{polynomial}]$ for $c=2,\dots, N$. In analogy with the analysis of the convergence for the $\sl(3)$ 
case we have to require $0<\phi_1-\phi_{c}<2\pi$ for $c=2,\dots, N$, which should be always possible to achieve without reducing the generality\footnote{With an exception for the boundary cases e.g. $\phi_c=0$, which can be obtained by taking the corresponding limits.}.

Plugging the explicit form of $\hat {\widebar O}$ from \eq{ObW} into \eq{weak3} we get
\beq\label{weakN}
\<Q_1\;\hat {\widebar O} \circ f\>_j 
=
\int_{-\infty}^\infty \mu_j(u)Q_1Q_\theta^- (1-R_1 {\cal D}^2)
\underbrace{
(1-R_2 {\cal D}^2)\dots
(1-R_N {\cal D}^2)
{\cal D}^{-N}  f}_{\equiv F(u)} \; du\;.
\eeq
Writing $R_1$ in an explicit way, and using the notation $F(u)$ for the product of all factors starting from the second acting on $f(u)$, we find
\beq\label{weakN2}
\<Q_1\;\hat {\widebar O} \circ f\>_j 
=
\int_{-\infty}^\infty \mu_j(u)\[Q_1(u)Q_\theta(u-\tfrac i2)  F(u)-Q_1(u+i)Q_\theta(u+\tfrac i2) F(u+i)\] \; du\;,
\eeq
Next we see that we can shift the integration contour for the second term down by $i$ to cancel precisely the first term.
Shifting the contour we have to be careful at $u=\theta_j-\tfrac i2$
where $\mu_j(u)$ has a simple pole. However, the factor $Q_\theta(u+\tfrac i2)$ vanishes exactly at $u=\theta_j-\tfrac i2$ ensuring that there are no extra contributions. There are no other poles to worry about because  $Q_1(u) F(u)$ is pole-free due to the Bethe ansatz equations, which can be seen via the same argument as for ${\cal W}$ itself before. This ends the proof of 
\eq{weak3} for general $\sl(N)$.

\subsection{Orthogonality relations}
Now having \eq{weak3} proven in the general case, we can simply repeat the same steps as in section \ref{sec:sl3}. Namely, instead of \eq{AB30}
for two  different Bethe states $|\Psi^A\>$ and $|\Psi^B\>$ we have
\beq\la{ABN0}
\langle Q_1^A\; ( \hat {\widebar O }^A - \hat {\widebar O}^B )\circ Q_{\bar{a}}^B \rangle_i = 0\;\;,\;\;a=1,\dots,N-1\;\;,\;\;i=1,\dots,L\;.
\eeq
Next we use again that the first and the last terms in $\hat {\widebar O}$ do not depend on the state by definition \eq{Bax2sl3} and we get
\beq\la{AB3gen}
\(\hat {\widebar O}^A-\hat {\widebar O}^B\)\circ Q_{\bar{a}}^B=
\sum_{j=1}^{L}\sum_{b=1}^{N-1}
(-1)^b(I_{b,j-1}^A-I_{b,j-1}^B)u^{j-1}\;{\cal D}^{[2b-N]}\circ Q_{\bar{a}}^B \;  .
\eeq
So it is clear that the generalization of \eq{detsl3} reads
\beq\la{mdef}
\<\Psi_A|\Psi_B\>\equiv \det_{(a,i),(b,j)}m_{(a,i),(b,j)}=0\;\;,\;\;
m_{(a,i),(b,j)}\equiv \left\langle Q_1^A \;u^{j-1}\; {\cal D}^{2b-N}\circ Q_{\bar{a}}^B \right\rangle_i
\;.
\eeq
for the case when the two states are different. We claim that this should give the orthogonality relation of two Bethe states written in SoV representation. 
Above we again use $(N-1)\times L$ dimensional multi-indexes $(a,i)$
and $(b,j)$ to indicate the determinant of the rectangular matrix of the dimension $(N-1)\times L$. Another form of this orthogonality relation is given in the introduction in  \eq{eq:Mhat}, \eq{M2}.

\subsection{Form factors}
In this section we generalise the considerations of section \ref{sec:FFs}, where
we introduced a particular type of form factors of the operators which can be obtained as a derivative of the integrals of motion w.r.t. some  
parameter $p$, which can be a twist angle $\phi_a,\;a=1,\dots,N-1$ or one of
 inhomogeneities $\theta_i,\;i=1,\dots,L$.
In section \ref{sec:FFs}, we considered $p=\phi_a$.  In general for both $p=\phi_a$ and $p=\theta_i$ we create quite a broad class of operators acting on the spin chain states, in total one can estimate that
$p=\theta_i$ creates $\sim (N-1)\times L^2$ operators $\d_{\theta_i}\hat I_{a,j-1}$ and for 
$p=\phi_a$ we get $\sim (N-1)^2\times L$ operators
$\d_{\phi_b}\hat I_{a,j-1}$. It is not immediately clear if all of them are independent and if they form a complete enough algebra of operators, so that the general spin chain operator can be obtained as a multiple action of those. We postpone these interesting questions to future studies.

In analogy with \eq{qdo} we have
\beq\la{ABN0v2}
\langle Q_1\;\d_p \hat {\widebar O }\circ Q_{\bar{a}} \rangle_i = 0\;\;,\;\;a=1,\dots,N-1\;\;,\;\;i=1,\dots,L\;.
\eeq
Note that the right way to generalize \eq{qdo} is to use
$\d_p \hat {\widebar O }$, rather than $\d_p  {\hat O }$, 
for exactly the same reason as in the previous section since \eq{weak3} discriminates between the two. We have
\beq
\d_p \hat {\widebar O } = \sum_{(b,j)}\d_p I_{b,j-1} u^{j-1}{\cal D}^{2b-N}+
\underbrace{\[\d_p Q_\theta^- {\cal D}^{-N}+(-1)^N\d_p Q_\theta^+ {\cal D}^N\]
+\sum_{b}\d_p I_{b,L} u^{L}{\cal D}^{2b-N}}_{\equiv -\hat Y_p} \;,
\eeq
where we denoted by $\hat{ Y}_p$ the inhomogeneous part of the linear system for $\d_p I_{n,j-1}$. Plugging into \eq{ABN0} we get
\beq\la{ABN1}
\sum_{(b,j)}m_{(a,i),(b,j)}\d_p I_{b,j-1} =
y_{(a,i)}\;\;,\;\;y_{(a,i)}\equiv \langle Q_1\;\hat Y_p\circ Q_{\bar{a}} \;, \rangle_i\;,
\eeq
where $m_{(a,i),(b,j)}$ is the same matrix as defined in the previous section in \eq{mdef} with two states taken to be the same.

Solving this system with Cramer's method we obtain the following structure
\beq
\d_p I_{c,k-1}=\frac{\det_{(a,i),(b,j)}\tilde m_{(a,i),(b,j)}}{\det_{(a,i),(b,j)}m_{(a,i),(b,j)}} \;,
\eeq
where $\tilde m_{(a,i),(b,j)}$ is the matrix
$m_{(a,i),(b,j)}$ with the column $(c,k)$ replaced with $y_{(a,i)}$ defined in \eq{ABN1}. Notice that the denominator has the meaning of the norm square $||\Psi||^2$ when comparing with \eq{mdef}.
Furthermore, both numerator and denominator can be written in the SoV-like form 
\beq
\d_p I_{c,k-1}=\frac{\int d{\bf x}\;\Psi^\dagger ({\bf x})\; \widehat {\widetilde M}({\bf x})\circ \Psi (\bf x)}
{\int d{\bf x}\;\Psi^\dagger ({\bf x}) \;\widehat M({\bf x})\circ \Psi ({\bf x})}\;,
\eeq
where we denote $\Psi ({\bf x})=\prod\limits_{(a,i)} Q_{\bar a}(x_{a,i})$
and
$\Psi^\dagger ({\bf x})=\prod\limits_{(a,i)} Q_{1}(x_{a,i})$\;.

\section{Conclusions}\label{sec:conclusions}

In this paper we have proposed the way to compute scalar products and form factors in SoV basis. Our method bypasses successfully the explicit construction of the separated variables and is valid for higher rank $\sl(N)$ spin chains.
Nevertheless, we hope that our result gives very strong hints of how to proceed with the first principle SoV construction too. 
We believe our construction should open the way to various new applications of the SoV methods beyond rank one systems. Let us discuss several of the promising future directions.

One of the important hints our result gives is that there should exist a ``dual'' SoV basis, potentially associated with some kind of $\hat C^{\rm good}$ operator, in analogy with the  $\mathfrak{su}(2)$ case. In this dual basis the wave function should factorise into the product of dual Q-functions, or Baxter polynomials at the last nesting level $Q_{\bar a}$. This observation of our paper could also resolve the problems outlined at the classical level in \cite{Martin:2015eea}. 
Another question is to build an explicit map from the natural spin chain variables to the separated variables like it was done for $\sl(2)$ in \cite{Derkachov:2002tf}. Having some explicit matrix elements, like those computed in this paper, could help to find an explicit integral transformation between these two bases. 

We derived an expression for the SoV type of scalar product of two Bethe states. It would be interesting to see if this expression remains the same when one of the states is taken off-shell (even though this may not be always well defined). 
A more well-posed problem is to relate our result with the Gaudin norm. We expect that they coincide up to a simple prefactor, and we expect the proof to go the same way as in Appendix of \cite{Gromov:2016itr} for the case of $\mathfrak{su}(2)$.
In regards to our results for the form-factors, it would be interesting to see if they could be generalised to the case with two different Bethe states.  

As a natural extension, it would be interesting to generalize our results to other types of spin chains based on $B_n,C_n,D_n$ Lie algebras, and especially to the supersymmetric case (particularly relevant for AdS/CFT applications), and also to various deformations, including trigonometric or even elliptic models, Gaudin models and boundary problems. It would be interesting to explore the implications of this construction for various classical/quantum and spectral dualities between integrable models \cite{Mironov:2012ba,Gorsky:2013xba,MTV1}.

While we have considered spin chains in a simple  infinite-dimensional highest-weight representation, we expect the results should generalize to other representations since we only use the Baxter equations which are quite universal. We have already explored several examples \cite{preparation} where  the same approach works for more involved principal series representations appearing in integrable fishnet CFTs, where it is important to also add a spacetime twist serving as a regulator \cite{TwistingFishing}. One of the methods one could try to use here is the {\it fusion}~\cite{Lipan:1997bs}, which should allow one to directly generalise any construction from the fundamental to any representation obtained as a tensor product of fundamentals.

Our results should also play a role in developing the SoV solution of the integrable fishnet CFT \cite{Gurdogan:2015csr} and of the {\it fishchain} model that serves as its dual \cite{Gromov:2019aku} and is reminiscent of Toda chains. The advantage of the fishchain model is that we can also analyse the SoV construction in the simplified settings of the classical regime.

In this paper we also considered a particular type of form factors of the operators which can be obtained as a variation of the integrals of motions w.r.t. some parameters. These form factors are analogous to the 3-point correlators of the type $\langle{\cal O}{\cal O}{\cal L}\rangle$ in the  fishnet theory or $\cN=4$ super Yang-Mills, where ${\cal L}$ is a  marginal operator such as the Lagrangian insertion\footnote{recently also considered in \cite{Basso:2018cvy}.} and ${\cal O}$ is a non-trivial single trace operator. SoV-type expressions for such structure constants, and even more general ones, have already been found in different parameter limits of $\cN=4$ SYM in a growing number of cases \cite{Cavaglia:2018lxi,Giombi:2018qox,Derkachov:2018rot,Giombi:2018hsx,preparation}\footnote{see also the recent results of \cite{Jiang:2019xdz} for a 3-point function expressed in terms of TBA solutions, suggesting additional fruitful connections.}, giving strong indications of the viability of a SoV strategy for correlators.  Generalization of our construction should give a closed totally non-perturbative expression for such 3-point correlator in terms of Q-functions, which are known from the Quantum Spectral Curve method developed in~\cite{Gromov:2013pga,Gromov:2014caa}, see \cite{Gromov:2017blm,Kazakov:2018hrh} for reviews. The simpler fishnet model should serve as an ideal playground to work out the details of the construction before uplifting it to the parent $\cN=4$ SYM theory. For the full $\cN=4$ SYM our results already suggest what structures to anticipate, for example we can expect to have the $\bQ_i$ and $\bQ^i$ Q-functions coupled in the scalar product.

\section*{Acknowledgements}

We are grateful to  D.~Grabner,  J.~Julius, V.~Kazakov,   A.~Sever, F.~Smirnov for related discussions. 
N.G. is also grateful to N.~Kitanine for inspiring discussions.
F.L.-M. also thanks for discussions S.~Derkachov, B.~Feigin,  G.~Ferrando, G.~Korchemsky, A.~Liashyk, D.~Serban and D.~Volin. A.C. thanks R.~Conti, S.~Negro and  R.~Tateo for fruitful discussions. This work is supported by Agence Nationale de la Recherche LabEx grant ENS-ICFP ANR-10-LABX-0010/ANR-10-IDEX-0001-02 PSL, and by the STFC grant (ST/P000258/1). 

\appendix

\section{Explicit result for the $\sl (3)$ scalar product at length 1}

\label{app:ex}

For the simplest higher rank example, namely the $\sl(3)$ inhomogenous spin chain with 1 site and twisted boundary conditions, our scalar product \eq{eq:Mhat} can be written in a compact determinant form 
\beq
    \langle\Psi_A|\Psi_B\rangle \;\propto\; \begin{vmatrix}
    \langle Q_1^A Q_{12}^{B+}\rangle & \langle Q_1^A Q_{12}^{B-}\rangle \\
    \langle Q_1^A Q_{13}^{B+}\rangle & \langle Q_1^A Q_{13}^{B-}\rangle \\
    \end{vmatrix}\;, 
\eeq
where
\beq
    \langle f\rangle = \int\limits_{-\infty}^{+\infty}dx\;\frac{1}{1+e^{2\pi (x-\theta_1)}} f(x)\;.
\eeq


\begin{thebibliography}{99}
  
  








  

\bibitem{Sklyanin:1984sb} 
  E.~K.~Sklyanin,
  ``The Quantum Toda Chain,''
  Lect.\ Notes Phys.\  {\bf 226}, 196 (1985).


\bibitem{Sklyanin:1987ih}
  E.~K.~Sklyanin,
  ``Separation of variables in the Gaudin model,''
  J.\ Sov.\ Math.\  {\bf 47} (1989) 2473
   [Zap.\ Nauchn.\ Semin.\  {\bf 164} (1987) 151].
  doi:10.1007/BF01840429


\bibitem{Sklyanin:1991ss}
  E.~K.~Sklyanin,
  ``Quantum inverse scattering method. Selected topics,''
  In: Quantum Group and Quantum Integrable Systems: Nankai Lectures
  on Mathematical Physics : Nankai Institute of Mathematics, China 2-18 April
  1991 (World Scientific 1992), pp 63-97
  [hep-th/9211111].



\bibitem{Sklyanin:1995bm}
  E.~K.~Sklyanin,
  ``Separation of variables - new trends,''
  Prog.\ Theor.\ Phys.\ Suppl.\  {\bf 118} (1995) 35
  doi:10.1143/PTPS.118.35
  [solv-int/9504001].


\bibitem{Korepin:1982gg}
  V.~E.~Korepin,
  ``Calculation Of Norms Of Bethe Wave Functions,''
  Commun.\ Math.\ Phys.\  {\bf 86} (1982) 391.




\bibitem{Kazama:2013rya}
  Y.~Kazama, S.~Komatsu and T.~Nishimura,
  ``A new integral representation for the scalar products of Bethe states for the XXX spin chain,''
  JHEP {\bf 1309} (2013) 013
  doi:10.1007/JHEP09(2013)013
  [arXiv:1304.5011 [hep-th]].



\bibitem{Niccoli:2012ci}
  G.~Niccoli,
  ``Antiperiodic spin-1/2 XXZ quantum chains by separation of variables: Complete spectrum and form factors,''
  Nucl.\ Phys.\ B {\bf 870} (2013) 397
  doi:10.1016/j.nuclphysb.2013.01.017
  [arXiv:1205.4537 [math-ph]].



\bibitem{Levy-Bencheton:2015mia}
  D.~Levy-Bencheton, G.~Niccoli and V.~Terras,
  ``Antiperiodic dynamical 6-vertex model by separation of variables II: Functional equations and form factors,''
  J.\ Stat.\ Mech.\  {\bf 1603} (2016) no.3,  033110
  doi:10.1088/1742-5468/2016/03/033110
  [arXiv:1507.03404 [math-ph]].
	


	\bibitem{Niccoli:2014sfa}
  G.~Niccoli and V.~Terras,
  ``Antiperiodic XXZ chains with arbitrary spins: Complete eigenstate construction by functional equations in separation of variables,''
  Lett.\ Math.\ Phys.\  {\bf 105} (2015) no.7,  989
  doi:10.1007/s11005-015-0759-9
  [arXiv:1411.6488 [math-ph]].


\bibitem{Jiang:2015lda}
  Y.~Jiang, S.~Komatsu, I.~Kostov and D.~Serban,
  ``The hexagon in the mirror: the three-point function in the SoV representation,''
  J.\ Phys.\ A {\bf 49} (2016) no.17,  174007
  doi:10.1088/1751-8113/49/17/174007
  [arXiv:1506.09088 [hep-th]].




\bibitem{Kitanine:2016pvg}
  N.~Kitanine, J.~M.~Maillet, G.~Niccoli and V.~Terras,
  ``The open XXX spin chain in the SoV framework: scalar product of separate states,''
  arXiv:1606.06917 [math-ph].
	


	\bibitem{Kitanine:2015jna}
  N.~Kitanine, J.~M.~Maillet, G.~Niccoli and V.~Terras,
  ``On determinant representations of scalar products and form factors in the SoV approach: the XXX case,''
  J.\ Phys.\ A {\bf 49} (2016) no.10,  104002
  doi:10.1088/1751-8113/49/10/104002
  [arXiv:1506.02630 [math-ph]].

	

	\bibitem{Kitanine:2014swa}
  N.~Kitanine, J.-M.~Maillet and G.~Niccoli,
  ``Open spin chains with generic integrable boundaries: Baxter equation and Bethe ansatz completeness from separation of variables,''
  J.\ Stat.\ Mech.\  {\bf 1405} (2014) P05015
  doi:10.1088/1742-5468/2014/05/P05015
  [arXiv:1401.4901 [math-ph]].


\bibitem{Derkachov:2001yn}
  S.~E.~Derkachov, G.~P.~Korchemsky and A.~N.~Manashov,
  ``Noncompact Heisenberg spin magnets from high-energy QCD: 1. Baxter Q operator and separation of variables,''
  Nucl.\ Phys.\ B {\bf 617} (2001) 375
  doi:10.1016/S0550-3213(01)00457-6
  [hep-th/0107193].



\bibitem{Derkachov:2002wz}
  S.~E.~Derkachov, G.~P.~Korchemsky, J.~Kotanski and A.~N.~Manashov,
  ``Noncompact Heisenberg spin magnets from high-energy QCD. 2. Quantization conditions and energy spectrum,''
  Nucl.\ Phys.\ B {\bf 645} (2002) 237
  doi:10.1016/S0550-3213(02)00842-8
  [hep-th/0204124].


\bibitem{Niccoli:2012vq}
  G.~Niccoli,
  ``Form factors and complete spectrum of XXX antiperiodic higher spin chains by quantum separation of variables,''
  J.\ Math.\ Phys.\  {\bf 54} (2013) 053516
  doi:10.1063/1.4807078
  [arXiv:1206.2418 [math-ph]].

\bibitem{Gromov:2016itr} 
  N.~Gromov, F.~Levkovich-Maslyuk and G.~Sizov,
  ``New Construction of Eigenstates and Separation of Variables for SU(N) Quantum Spin Chains,''
  JHEP {\bf 1709}, 111 (2017)
  doi:10.1007/JHEP09(2017)111
  [arXiv:1610.08032 [hep-th]].
  


\bibitem{Martin:2015eea} 
  D.~Martin and F.~Smirnov,
  ``Problems with using separated variables for computing expectation values for higher ranks,''
  Lett.\ Math.\ Phys.\  {\bf 106}, no. 4, 469 (2016)
  doi:10.1007/s11005-016-0823-0
  [arXiv:1506.08042 [math-ph]].



\bibitem{Beisert:2010jr}
  N.~Beisert {\it et al.},
  ``Review of AdS/CFT Integrability: An Overview,''
  Lett.\ Math.\ Phys.\  {\bf 99} (2012) 3
  [arXiv:1012.3982 [hep-th]].


\bibitem{Sklyanin:1992eu}
  E.~K.~Sklyanin,
  ``Separation of variables in the classical integrable SL(3) magnetic chain,''
  Commun.\ Math.\ Phys.\  {\bf 150} (1992) 181
  doi:10.1007/BF02096572
  [hep-th/9211126].


\bibitem{Sklyanin:1992sm}
  E.~K.~Sklyanin,
  ``Separation of variables in the quantum integrable models related to the Yangian Y[sl(3)],''
  J.\ Math.\ Sci.\  {\bf 80} (1996) 1861
   [Zap.\ Nauchn.\ Semin.\  {\bf 205} (1993) 166]
  doi:10.1007/BF02362784
  [hep-th/9212076].




\bibitem{Smirnov2001}
F.~Smirnov, ``Separation of variables for quantum integrable
models related to $U_q(\widehat{sl}_N)$'', math-ph/0109013



\bibitem{Scott:1994dz}
  D.~R.~D.~Scott,
  ``Classical functional Bethe ansatz for SL(N): Separation of variables for the magnetic chain,''
  J.\ Math.\ Phys.\  {\bf 35} (1994) 5831
  doi:10.1063/1.530712
  [hep-th/9403030].

\bibitem{Ge95}	
	M.~I.~Gekhtman, Separation of variables in the classical SL(N) magnetic chain, Comm. Math.
Phys. Volume 167, Number 3 (1995), 593-605. http://projecteuclid.org/euclid.cmp/1104272160

  	\bibitem{Chervov:2006xk}
  A.~Chervov and D.~Talalaev,
  ``Quantum spectral curves, quantum integrable systems and the geometric Langlands correspondence,''
  hep-th/0604128.
  

\bibitem{Chervov:2007bb}
  A.~Chervov and G.~Falqui,
  ``Manin matrices and Talalaev's formula,''
  J.\ Phys.\ A {\bf 41} (2008) 194006
  doi:10.1088/1751-8113/41/19/194006
  [arXiv:0711.2236 [math.QA]].
  

  \bibitem{Kulish:1979cr} 
  P.~P.~Kulish and N.~Y.~Reshetikhin,
  ``Generalized Heisenberg Ferromagnet And The Gross-neveu Model,''
  Sov.\ Phys.\ JETP {\bf 53}, 108 (1981)
  [Zh.\ Eksp.\ Teor.\ Fiz.\  {\bf 80}, 214 (1981)].
  

  \bibitem{Kulish:1985bj} 
  P.~P.~Kulish,
  ``Integrable graded magnets,''
  J.\ Sov.\ Math.\  {\bf 35}, 2648 (1986)
  [Zap.\ Nauchn.\ Semin.\  {\bf 145}, 140 (1985)].
  

  \bibitem{Krichever:1996qd} 
  I.~Krichever, O.~Lipan, P.~Wiegmann and A.~Zabrodin,
  ``Quantum integrable systems and elliptic solutions of classical discrete nonlinear equations,''
  Commun.\ Math.\ Phys.\  {\bf 188}, 267 (1997)
  [hep-th/9604080].
  

\bibitem{Dorey:2006an} 
  P.~Dorey, C.~Dunning, D.~Masoero, J.~Suzuki and R.~Tateo,
  ``Pseudo-differential equations, and the Bethe ansatz for the classical Lie algebras,''
  Nucl.\ Phys.\ B {\bf 772}, 249 (2007)
  [hep-th/0612298].


  \bibitem{Kazakov:2007fy} 
  V.~Kazakov, A.~S.~Sorin and A.~Zabrodin,
  ``Supersymmetric Bethe ansatz and Baxter equations from discrete Hirota dynamics,''
  Nucl.\ Phys.\ B {\bf 790}, 345 (2008)
  [hep-th/0703147 [HEP-TH]].



\bibitem{Gromov:2007ky} 
  N.~Gromov and P.~Vieira,
  ``Complete 1-loop test of AdS/CFT,''
  JHEP {\bf 0804}, 046 (2008)
  doi:10.1088/1126-6708/2008/04/046
  [arXiv:0709.3487 [hep-th]].


 \bibitem{Kazakov:2015efa} 
  V.~Kazakov, S.~Leurent and D.~Volin,
  ``T-system on T-hook: Grassmannian Solution and Twisted Quantum Spectral Curve,''
  JHEP {\bf 1612}, 044 (2016)
  doi:10.1007/JHEP12(2016)044
  [arXiv:1510.02100 [hep-th]].
 

\bibitem{Gromov:2017blm}
  N.~Gromov,
  ``Introduction to the Spectrum of ${\cal N}=4$ SYM and the Quantum Spectral Curve,''
  arXiv:1708.03648 [hep-th].
  


\bibitem{Kazakov:2018hrh} 
  V.~Kazakov,
  ``Quantum Spectral Curve of $\gamma$-twisted $\mathcal{ N}=4$ SYM theory and fishnet CFT,''
  Rev.\ Math.\ Phys.\  {\bf 30}, no. 07, 1840010 (2018)
  [arXiv:1802.02160 [hep-th]].



  

 
 
 \bibitem{Sklyanin:1989cg}
  E.~K.~Sklyanin,
  ``New Approach To The Quantum Nonlinear Schrodinger Equation,'' (see also the Russian version of this paper)
  J.\ Phys.\ A {\bf 22} (1989) 3551.
  doi:10.1088/0305-4470/22/17/020. 
  

  \bibitem{Derkachov:2002tf} 
  S.~E.~Derkachov, G.~P.~Korchemsky and A.~N.~Manashov,
  ``Separation of variables for the quantum SL(2,R) spin chain,''
  JHEP {\bf 0307}, 047 (2003)
  doi:10.1088/1126-6708/2003/07/047
  [hep-th/0210216].
  

 
 










  \bibitem{Liashyk:2018qfc} 
  A.~Liashyk and N.~A.~Slavnov,
  ``On Bethe vectors in $\mathfrak{gl}_3$-invariant integrable models,''
  JHEP {\bf 1806}, 018 (2018)
  doi:10.1007/JHEP06(2018)018
  [arXiv:1803.07628 [math-ph]].
  
 

\bibitem{Ryan:2018fyo} 
  P.~Ryan and D.~Volin,
  ``Separated variables and wave functions for rational gl(N) spin chains in the companion twist frame,''
  J.\ Math.\ Phys.\  {\bf 60}, no. 3, 032701 (2019)
  doi:10.1063/1.5085387
  [arXiv:1810.10996 [math-ph]].
  
  
  
  
  
 

 \bibitem{Derkachov:2018ewi} 
  S.~E.~Derkachov and P.~A.~Valinevich,
  ``Separation of variables for the quantum $SL(3,\mathbb C)$ spin magnet: eigenfunctions of Sklyanin $B$-operator,''
  Zap.\ Nauchn.\ Semin.\  {\bf 473}, 110 (2018)
  [arXiv:1807.00302 [math-ph]].
  
  
  


\bibitem{Maillet:2018bim} 
  J.~M.~Maillet and G.~Niccoli,
  ``On quantum separation of variables,''
  J.\ Math.\ Phys.\  {\bf 59}, no. 9, 091417 (2018)
  doi:10.1063/1.5050989
  [arXiv:1807.11572 [math-ph]].
 
  
  \bibitem{Maillet:2018czd}
  J.~M.~Maillet and G.~Niccoli,
  ``Complete spectrum of quantum integrable lattice models associated to Y(gl(n)) by separation of variables,''
  SciPost Phys.\  {\bf 6} (2019) 071
  doi:10.21468/SciPostPhys.6.6.071
  [arXiv:1810.11885 [math-ph]].
  
  \bibitem{Maillet:2018rto}
  J.~M.~Maillet and G.~Niccoli,
  ``Complete spectrum of quantum integrable lattice models associated to $\mathcal{U}_{q} (\widehat{gl_{n}})$ by separation of variables,''
  arXiv:1811.08405 [math-ph].
  

\bibitem{Derkachov:2010qe}
  S.~E.~Derkachov and A.~N.~Manashov,
  ``Noncompact sl(N) spin chains: BGG-resolution, Q-operators and alternating sum representation for finite dimensional transfer matrices,''
  Lett.\ Math.\ Phys.\  {\bf 97} (2011) 185
  doi:10.1007/s11005-011-0472-2
  [arXiv:1008.4734 [nlin.SI]].




\bibitem{Derkachov:2006fz}
  S.~E.~Derkachov and A.~N.~Manashov,
  ``Baxter operators for the quantum sl(3) invariant spin chain,''
  J.\ Phys.\ A {\bf 39} (2006) 13171
  doi:10.1088/0305-4470/39/42/001
  [nlin/0604018 [nlin-si]].
  

\bibitem{Derkachov:2005hw}
  S.~E.~Derkachov,
  ``Factorization of the R-matrix. I.,''
  math/0503396 [math-qa].



\bibitem{Gromov:2018cvh} 
  N.~Gromov and F.~Levkovich-Maslyuk,
  ``New Compact Construction of Eigenstates for Supersymmetric Spin Chains,''
  JHEP {\bf 1809}, 085 (2018)
  doi:10.1007/JHEP09(2018)085
  [arXiv:1805.03927 [hep-th]].

\bibitem{Gurdogan:2015csr} 
  Ö.~Gürdoğan and V.~Kazakov,
  ``New Integrable 4D Quantum Field Theories from Strongly Deformed Planar $\mathcal N = $ 4 Supersymmetric Yang-Mills Theory,''
  Phys.\ Rev.\ Lett.\  {\bf 117}, no. 20, 201602 (2016)
  Addendum: [Phys.\ Rev.\ Lett.\  {\bf 117}, no. 25, 259903 (2016)]
  doi:10.1103/PhysRevLett.117.201602, 10.1103/PhysRevLett.117.259903
  [arXiv:1512.06704 [hep-th]].


\bibitem{Costa:2010rz} M.~S.~Costa, R.~Monteiro, J.~E.~Santos and D.~Zoakos, ``On three-point correlation functions in the gauge/gravity duality,''
  JHEP {\bf 1011} (2010) 141
  doi:10.1007/JHEP11(2010)141
  [arXiv:1008.1070 [hep-th]].

\bibitem{Cavaglia:2018lxi} 
  A.~Cavagli\`a, N.~Gromov and F.~Levkovich-Maslyuk,
  ``Quantum spectral curve and structure constants in $ \mathcal{N}=4 $ SYM: cusps in the ladder limit,''
  JHEP {\bf 1810}, 060 (2018)
  doi:10.1007/JHEP10(2018)060
  [arXiv:1802.04237 [hep-th]].
  

  \bibitem{preparation} 
  A.~Cavagli\`a, N.~Gromov, F.~Levkovich-Maslyuk and A.~Sever,
  to appear


\bibitem{SmirnovQuantM}
F.~Smirnov, V.~Zeitlin,
``On The Quantization of Affine Jacobi Varieties of Spectral Curves''. Statistical Field Theories. – Springer, Dordrecht, 2002, p. 79-89.


\bibitem{SmirnovClassM}
F.~Smirnov, V.~Zeitlin,
``Affine Jacobians of spectral curves and integrable models'', arXiv:math-ph/0203037


\bibitem{Gromov:2019aku} 
  N.~Gromov and A.~Sever,
  ``The Holographic Fishchain,''
  arXiv:1903.10508 [hep-th].

  \bibitem{Lipan:1997bs} 
  O.~Lipan, P.~B.~Wiegmann and A.~Zabrodin,
  ``Fusion rules for quantum transfer matrices as a dynamical system on Grassmann manifolds,''
  Mod.\ Phys.\ Lett.\ A {\bf 12}, 1369 (1997)
  doi:10.1142/S0217732397001394
  [solv-int/9704015].
  


\bibitem{SutherlandN1}
B.~Sutherland, ``A General Model for Multicomponent Quantum Systems,'' Phys.
Rev. {\bf B12} (1975) 3795-3805.


\bibitem{Kulish:1983rd}
  P.~P.~Kulish and N.~Y.~Reshetikhin,
  ``Diagonalization Of Gl(n) Invariant Transfer Matrices And Quantum N Wave System (lee Model),''
  J.\ Phys.\ A {\bf 16} (1983) L591.
  doi:10.1088/0305-4470/16/16/001
  


\bibitem{Lukyanov:2000jp} 
  S.~L.~Lukyanov,
  ``Finite temperature expectation values of local fields in the sinh-Gordon model,''
  Nucl.\ Phys.\ B {\bf 612}, 391 (2001)
  doi:10.1016/S0550-3213(01)00365-0
  [hep-th/0005027].
 
 
 \bibitem{Bazhanov:1996dr}
  V.~V.~Bazhanov, S.~L.~Lukyanov and A.~B.~Zamolodchikov,
  ``Integrable structure of conformal field theory. 2. Q operator and DDV equation,''
  Commun.\ Math.\ Phys.\  {\bf 190} (1997) 247
  doi:10.1007/s002200050240
  [hep-th/9604044].

\bibitem{Pronko:1999gh} 
  G.~P.~Pronko and Y.~G.~Stroganov,
  ``The Complex of solutions of the nested Bethe ansatz. The A(2) spin chain,''
  J.\ Phys.\ A {\bf 33}, 8267 (2000)
  doi:10.1088/0305-4470/33/46/309
  [hep-th/9902085].


\bibitem{Gromov:2010kf} 
  N.~Gromov and V.~Kazakov,
  ``Review of AdS/CFT Integrability, Chapter III.7: Hirota Dynamics for Quantum Integrability,''
  Lett.\ Math.\ Phys.\  {\bf 99}, 321 (2012)
  doi:10.1007/s11005-011-0513-x
  [arXiv:1012.3996 [hep-th]].
  

  	\bibitem{Mironov:2012ba}
  A.~Mironov, A.~Morozov, B.~Runov, Y.~Zenkevich and A.~Zotov,
  ``Spectral Duality Between Heisenberg Chain and Gaudin Model,''
  Lett.\ Math.\ Phys.\  {\bf 103} (2013) no.3,  299
  doi:10.1007/s11005-012-0595-0
  [arXiv:1206.6349 [hep-th]].
	


	\bibitem{Gorsky:2013xba}
  A.~Gorsky, A.~Zabrodin and A.~Zotov,
  ``Spectrum of Quantum Transfer Matrices via Classical Many-Body Systems,''
  JHEP {\bf 1401} (2014) 070
  doi:10.1007/JHEP01(2014)070
  [arXiv:1310.6958 [hep-th]].
	


	\bibitem{MTV1}
	E.~Mukhin, V.~Tarasov, A.~Varchenko, ``Bispectral and $(gl_N,gl_M)$ dualities'', [math.QA/0510364]; $\bullet$ E.~Mukhin, V.~Tarasov, A.~Varchenko, ``Bispectral and $(gl_N,gl_M)$ dualities, discrete versus differential'', Advances in Mathematics, 218 (2008)
216-265, math.QA/0605172.




    \bibitem{TwistingFishing} 
  A.~Cavagli\`a, N.~Grabner, N.~Gromov and A.~Sever,
  ``Twisting and Fishing,''
  in preparation.

\bibitem{Basso:2018cvy} 
  B.~Basso, J.~Caetano and T.~Fleury,
  ``Hexagons and Correlators in the Fishnet Theory,''
  arXiv:1812.09794 [hep-th].
  

  \bibitem{Giombi:2018qox} 
  S.~Giombi and S.~Komatsu,
  ``Exact Correlators on the Wilson Loop in $\mathcal{N}=4$ SYM: Localization, Defect CFT, and Integrability,''
  JHEP {\bf 1805}, 109 (2018)
  [arXiv:1802.05201 [hep-th]].
  

\bibitem{Derkachov:2018rot} 
  S.~Derkachov, V.~Kazakov and E.~Olivucci,
  ``Basso-Dixon Correlators in Two-Dimensional Fishnet CFT,''
  JHEP {\bf 1904}, 032 (2019)
  doi:10.1007/JHEP04(2019)032
  [arXiv:1811.10623 [hep-th]].

  

  \bibitem{Giombi:2018hsx} 
  S.~Giombi and S.~Komatsu,
  ``More Exact Results in the Wilson Loop Defect CFT: Bulk-Defect OPE, Nonplanar Corrections and Quantum Spectral Curve,''
  J.\ Phys.\ A {\bf 52}, no. 12, 125401 (2019)
  doi:10.1088/1751-8121/ab046c
  [arXiv:1811.02369 [hep-th]].
  

\bibitem{Jiang:2019xdz} 
  Y.~Jiang, S.~Komatsu and E.~Vescovi,
  ``Structure Constants in $\mathcal{N}=4$ SYM at Finite Coupling as Worldsheet $g$-Function,''
  arXiv:1906.07733 [hep-th].

\bibitem{Gromov:2013pga}
  N.~Gromov, V.~Kazakov, S.~Leurent and D.~Volin,
  ``Quantum Spectral Curve for Planar $\mathcal{N} =4$ Super-Yang-Mills Theory,''
  Phys.\ Rev.\ Lett.\  {\bf 112} (2014) no.1,  011602
  doi:10.1103/PhysRevLett.112.011602
  [arXiv:1305.1939 [hep-th]].


\bibitem{Gromov:2014caa}
  N.~Gromov, V.~Kazakov, S.~Leurent and D.~Volin,
  ``Quantum spectral curve for arbitrary state/operator in AdS$_{5}$/CFT$_{4}$,''
  JHEP {\bf 1509} (2015) 187
  doi:10.1007/JHEP09(2015)187
  [arXiv:1405.4857 [hep-th]].

  
  

\end{thebibliography}
\end{document}